\documentclass[journal]{IEEEtran}

\usepackage{amsmath}
\usepackage{subfigure}
\usepackage{lineno,hyperref}
\usepackage{xcolor}
\usepackage{extarrows}
\usepackage{algorithm}
\usepackage{algorithmic}
\usepackage{amssymb}
\usepackage{bm}
\usepackage{graphicx}
\usepackage[justification=centering]{caption}
\usepackage{epstopdf}

\newcommand{\ba}{\boldsymbol{a}}

\newcommand{\bb}{\boldsymbol{b}}
\newcommand{\bc}{\boldsymbol{c}}
\newcommand{\bx}{\boldsymbol{x}}
\newcommand{\by}{\boldsymbol{y}}

\newcommand{\bw}{\boldsymbol{w}}
\newcommand{\bz}{\boldsymbol{z}}
\newcommand{\bp}{\boldsymbol{p}}
\newcommand{\bs}{\boldsymbol{s}}
\newcommand{\bq}{\boldsymbol{q}}
\newcommand{\br}{\boldsymbol{r}}

\newcommand{\bX}{\boldsymbol{X}}

\newcommand{\bY}{\boldsymbol{Y}}
\newcommand{\bA}{\boldsymbol{A}}
\newcommand{\bB}{\boldsymbol{B}}
\newcommand{\bR}{\boldsymbol{R}}
\newcommand{\bZ}{\boldsymbol{Z}}
\newcommand{\bW}{\boldsymbol{W}}
\newcommand{\bI}{\boldsymbol{I}}

\newcommand{\bC}{\boldsymbol{C}}
\newcommand{\bU}{\boldsymbol{U}}
\newcommand{\bV}{\boldsymbol{V}}

\newcommand{\bLambda}{\boldsymbol{\Lambda}}

\newcommand{\N}{\mathcal{N}}

\newcommand{\bnu}{\boldsymbol{\nu}}
\newcommand{\cev}[1]{\reflectbox{\ensuremath{\vec{\reflectbox{\ensuremath{#1}}}}}}

\newcommand{\U}{\boldsymbol{U}}

\newcommand{\bPhi}{\boldsymbol{\Phi}}

\newcommand{\bomega}{\boldsymbol{\omega}}

\newcommand{\bphi}{\boldsymbol{\phi}}

\newcommand{\rv}{\textcolor{black}}
\newcommand{\rev}{\textcolor{black}}

\begin{document}

\title{Approximate Message Passing with Unitary Transformation for Robust Bilinear Recovery}
\author{Zhengdao~Yuan, Qinghua Guo, \IEEEmembership{Senior Member,~IEEE}, and Man Luo
\thanks{Z. Yuan's work was supported by the National Natural Science
Foundation of China (61901417, 61571402), Postdoctoral science foundation of China (2019M652576), Henan research project of high education (20B510005), and Science and technology research project of Henan province (202102210313, 202102210172).}
\thanks{Corresponding author: Q. Guo (qguo@uow.edu.au).}
\thanks{Z. Yuan is with the Artificial Intelligence Technology Engineering Research Center, Open University of Henan, and School of Information Engineering, Zhengzhou University, Zhengzhou 450002, China. He was also with the School of Electrical, Computer and Telecommunications Engineering, University of Wollongong, Wollongong, NSW 2522, Australia (e-mail: yuan\_zhengdao@163.com).}
\thanks{Q. Guo and M. Luo are with the School of Electrical, Computer and Telecommunications Engineering, University of Wollongong, Wollongong, NSW 2522, Australia  (e-mail: qguo@uow.edu.au, ml857@uowmail.edu.au).}
\thanks{Accepted by IEEE Transactions on Signal Processing.}
}
\markboth{AMP with Unitary Transformation for Bilinear Recovery}%
{Shell \MakeLowercase{\textit{et al.}}: Bare Demo of IEEEtran.cls for IEEE Journals}

\maketitle

\begin{abstract}
Recently, several promising approximate message passing (AMP) based algorithms have been developed for bilinear recovery with model $\bY=\sum_{k=1}^K b_k \bA_k \bC +\bW $, where $\{b_k\}$ and $\bC$ are jointly recovered with known $\bA_k$ from the noisy measurements $\bY$. The bilinear recovery problem has many applications such as dictionary learning, self-calibration, compressive sensing with matrix uncertainty, etc. In this work, we propose a new \rv{approximate Bayesian inference algorithm for bilinear recovery, where AMP with unitary transformation (UTAMP) is integrated with belief propagation (BP), variational inference (VI) and expectation propagation (EP) to achieve efficient approximate inference}. It is shown that, compared to state-of-the-art \rv{bilinear recovery} algorithms, the proposed algorithm is much more robust and faster, leading to remarkably better performance.
\end{abstract}


\section{Introduction}
In this work, we consider the following bilinear problem
\begin{eqnarray}
\bY=\sum_{k=1}^K b_k\bA_k\bC+\bW, \label{eq:intro1}
\end{eqnarray}
where $\bY$ denotes measurements, matrices $\{\bA_k\}$ are known, $\{b_k\}$ and $\bC$ are to be recovered, and $\bW$ represents white Gaussian noise. When $\bY$, $\bC$ \rv{and $\bW$}  are replaced with the \rv{corresponding} vectors $\by$, $\bc$ \rv{and $\bw$}, respectively, the above multiple measurement vector (MMV) problem is reduced to a single measurement vector (SMV) problem. Model (\ref{eq:intro1}) covers a variety of problems, e.g., compressive sensing (CS) with matrix uncertainty \cite{Zhu2011CSMU}, joint channel estimation and detection \cite{Ghassan1994}, self-calibration and blind deconvolution \cite{Ling2015}, and structured dictionary learning \cite{Rubinstein2010DL}.


Recently, several approximate message passing (AMP) \cite{Donoho2010a} \cite{Donoho2010b} based algorithms have been developed to solve the bilinear problem, which show promising performance, compared to existing non-message passing alternates \cite{Sarkar2019}. The generalized AMP (GAMP) \cite{GAMP2011} was extended to bilinear GAMP (BiGAMP) \cite{Parker2014I} for solving a general bilinear problem, i.e., recover both $\bA$ and $\bX$ from observation $\bY=\bA\bX+\bW$. The parametric BiGAMP (P-BiGAMP) is then proposed in \cite{Parker2016}, which works with model \eqref{eq:intro1} to jointly recover $\{b_k\}$ and $\bC$. Lifted AMP was proposed in \cite{Davenport2016} by using the lifting approach \cite{Ahmed2014}, \cite{Ling_2015}. However, these AMP based algorithms are vulnerable to difficult $\bA$  matrices, e.g., ill-conditioned, correlated, rank-deficient or non-zero mean matrices as AMP can easily diverge in these cases \cite{rangana2019convergence}.

It was discovered in \cite{Guo2015UtAMP} that the AMP algorithm can still perform well for difficult  $\bA$. Instead of working directly with the original model $\by=\bA\bx+\bw$, \cite{Guo2015UtAMP} proposed to apply AMP to a unitary transform of the original model, i.e., $\bU^H\by=\bLambda\bV\bx+\U^H\bw$ where the unitary matrix $\bU$ can be obtained by the singular value decomposition (SVD) of matrix $\bA$, i.e., $\bA=\bU\bLambda\bV$. \rv{In the case of circulant matrix $\bA$, the matrix $\bU^H$ for unitary transformation can be simply the normalized discrete Fourier transform matrix, which allows fast implementation with the fast Fourier transform (FFT) algorithm.} AMP with unitary transformation, called UTAMP for convenience, was inspired by the work in \cite{guo2013}, which can be regarded as the first application of UTAMP to turbo equalization, where the normalized discrete Fourier transform matrix is used for unitary transformation. UTAMP was also recently employed for sparse Bayesian learning (SBL) \cite{UTAMPSBL2019}, \rv{which} shows outstanding performance even with difficult measurement matrices. The application of UTAMP to inverse synthetic aperture radar (ISAR) imaging was also studied in \cite{Radar2020}, and real data experiments show its excellent capability of achieving high Doppler resolution with low complexity, where the measurement matrix can be highly correlated to achieve high Doppler resolution. \rv{UTAMP has also been employed for low complexity direction of arrival (DOA) estimation \cite{DOAUT} and iterative detection for orthogonal time frequency space modulation (OTFS) \cite{OTFSUT}, which shows promising performance.} These motivated us to design efficient and robust bilinear recovery algorithms with UTAMP in this work.

Most recently, to achieve robust bilinear recovery, building on vector AMP (VAMP) \cite{rangan2019}, the lifted VAMP was proposed in \cite{Fletcher2018}, and the bilinear adaptive VAMP (BAd-VAMP) was proposed in \cite{Sarkar2019}, which inherit the robustness of VAMP. It was shown that BAd-VAMP is more robust and faster, and it can outperform lifted VAMP significantly \cite{Sarkar2019}. \rv{Based on VAMP, PC-VAMP was proposed in \cite{ZHU} to achieve compressive sensing with structured matrix perturbation. In \cite{MengZhu}, BAd-VAMP was extended to incorporate arbitrary distributions on the output transform based on the framework in \cite{XiangMing}.}

\begin{algorithm}
	\caption{Vector Stepsize AMP}
	Initialize $\bm{\tau}_x^{(0)}>0$ and $\bx^{(0)}$. Set $\bs^{(-1)}=\bm{0}$ and $ t=0$.\\
	\textbf{Repeat}
	\begin{algorithmic}[1]
		\STATE $\bm{\tau}_p$ = $ | \bA |^2  \bm{\tau}^t_x$\\
		\STATE $\bp= \bA \bx^t - \bm{\tau}_{p}  \cdot \bs^{t-1} $\\
		\STATE $\bm{\tau}_s = \mathbf{1} ./ (\bm{\tau}_p+\beta^{-1} \mathbf{1}) $\\
		\STATE $ \bs^t=\bm{\tau}_s \cdot  (\by-\bp) $\\
		
		\STATE 	$ \mathbf{1} ./\bm{\tau}_q =  | \bA^H |^2   \bm{\tau}_s  $\\
		\STATE $ \bq = \bx^t + \bm{\tau}_q \cdot  \bA^H   \bs^t $\\
		\STATE 	$\bm{\tau}_x^{t+1}$ = $ \bm{\tau}_q  \cdot   g_{x}' ( \bq, \bm{\tau}_q)  $\\
		\STATE 	$\mathbf{x}^{t+1} = g_{x}  ( \bq, \bm{\tau}_q)$	
		\STATE $t=t+1$
	\end{algorithmic}
	\textbf{Until terminated}
	\label{the vector stepsize AMP-table}
\end{algorithm}

\begin{algorithm}
	\caption{UTAMP Version 1}
	Unitary transform: $\br=\bU^H \by =\bPhi \bx+\bm{\omega}$, where $\bPhi=\bU^H\bA=\bLambda \bV$, and $\bU$ is obtained from the SVD $\bA=\bU \bLambda\bV$.\\
	Initialize $\bm{\tau}_x^{(0)}>0$ and $\bx^{(0)}$. Set $\bs^{(-1)}=\bm{0}$ and $ t=0$.\\
	\textbf{Repeat}
	\begin{algorithmic}[1]
		\STATE $\bm{\tau}_p$ = $ | \mathbf{\Phi} |^2  \bm{\tau}^t_x$\\
		\STATE $\bp= \mathbf{ \mathbf{\Phi} x^t} - \bm{\tau}_{p}  \cdot  \bs^{t-1} $\\
		\STATE $\bm{\tau}_s = \mathbf{1} ./ (\bm{\tau}_p+\beta^{-1} \mathbf{1}) $\\
		\STATE $ \bs^t=\bm{\tau}_s \cdot  (\br-\bp) $\\
		
		\STATE 	$ \mathbf{1} ./\bm{\tau}_q =  | \mathbf{\Phi}^H |^2   \bm{\tau}_s  $\\
		\STATE $ \bq = \bx^t + \bm{\tau}_q \cdot  (\mathbf{\Phi}^H \bs^t) $\\
		\STATE 	$\bm{\tau}_x^{t+1}$ = $ \bm{\tau}_q  \cdot   g_{x}' ( \bq, \bm{\tau}_q)  $\\
		\STATE 	$\bx^{t+1} = g_{x}  ( \bq, \bm{\tau}_q)$	
		\STATE $t=t+1$
	\end{algorithmic}
	\textbf{Until terminated}
	\label{UTAMPv1}
\end{algorithm}

In this work, \rv{leveraging UTAMP, we propose} a more robust and faster approximate inference algorithm for bilinear recovery, which is called Bi-UTAMP. By using the lifting approach, the original bilinear problem is reformulated as a linear one. Then, \rv{the structured variational inference (VI) \cite{VMP1}, \cite{VMPnew}, \cite{VMPJustin}, expectation propagation (EP) \cite{EP} and belief propagation (BP) \cite{BPa}, \cite{Kschischang2001} are combined with UTAMP, where UTAMP is employed to handle the most computational intensive part, leading to the fast and robust approximate inference algorithm Bi-UTAMP.} It is shown that Bi-UTAMP performs significantly better and is much faster than \rv{state-of-the-art bilinear recovery algorithms} for difficult matrices.

The remainder of this paper is organized as follows. In Section II, \rv{we briefly introduce the (UT)AMP algorithms, which form the basis for developing Bi-UTAMP}. Bi-UTAMP is designed for SMV problems and it is then extended for MMV problems in Section III. Numerical examples \rv{and comparisons with state-of-the-art message passing and non-message passing algorithms} are provided in Section IV, and conclusions are drawn in Section V.

\begin{algorithm}
	\caption{UTAMP Version 2}
	Unitary transform: $\br=\bU^H \by =\bPhi \bx+\bm{\omega}$, where $\bPhi=\bU^H\bA=\bLambda \bV$, and $\bU$ is obtained from the SVD $\bA=\bU \bLambda \bV$. \\
	Define vector $\bm{\lambda}=\mathbf{ \Lambda \Lambda}^H \textbf{1}$.\\
	Initialize ${\tau}_x^{(0)}>0$ and $\bx^{(0)}$. Set $\bs^{(-1)}=\bm{0}$ and $ t=0$.\\
	\textbf{Repeat}
	\begin{algorithmic}[1]
		
		\STATE $\quad		\bm{\tau}_p$ = $ \tau^t_x  \bm{\lambda}$\\
		\STATE $\quad		\bp= \mathbf{\Phi} \bx^t - \bm{\tau}_{p} \cdot  \bs^{t-1} $\\
		\STATE $\quad		\bm{\tau}_s = \mathbf{1}./ (\bm{\tau}_p+\beta^{-1} \mathbf{1}) $\\
		\STATE $\quad		\bs^t= \bm{\tau}_s \cdot (\br-\bp) $\\
		
		\STATE 	$\quad		1/\tau_q = ({1}/{N}) \bm{\lambda}^T \bm{\tau }_s   $\\
		\STATE $\quad		 \bq = \bx^t + \tau_q  \mathbf{\Phi}^H \bs^t$\\
		\STATE 	$\quad		\tau_x^{t+1}$ = $ (\tau_q  /N)  \mathbf{1}^H   g_{x}' ( \bq, \tau_q) $\\
		\STATE 	$\quad		\mathbf{x}^{t+1} = g_{x}  ( \bq, \tau_q)$	
		\STATE 	$\quad	      t=t+1$
	\end{algorithmic}
	\textbf{Until terminated}
	\label{UTAMP}
\end{algorithm}

\textit{Notations}- Boldface lower-case and upper-case letters denote vectors and matrices, respectively. Superscripts $(\cdot)^H$ and $(\cdot)^T$ represent conjugate transpose and transpose, respectively, and $(\cdot)^*$ represents the conjugate operation.
A Gaussian distribution of $x$ with mean {$\hat x$} and variance $\nu_x$ is represented by $\N(x;{\hat x},\nu_x)$. We also simply use $\N(m, v)$ to represent a Gaussian distribution with mean $m$ and variance $v$. Notation $\otimes$ represents the Kronecker product. The relation $f(x)=cg(x)$ for some positive constant $c$ is written as $f(x)\propto g(x)$. We use $\ba\cdot\bb$ and $\ba\cdot/\bb$ to represent the element-wise product and division between vectors $\ba$ and $\bb$, respectively. The notation $\ba^{.-1}$ denotes the element-wise inverse operation to vector $\ba$.
We use $|\bA|^{2}$ to denote element-wise magnitude squared operation for  $\bA$, and use $||\ba||^2$ to denote the squared $l_2$ norm of $\ba$.  The notation $<\ba>$ denotes the average operation for $\ba$, i.e., the sum of the elements of $\ba$ divided by its length. The notation $\int_{\bc\vee {c_n}} f_{\bc}(\bc)$ represents integral over all elements in $\bc$ except $c_n$.  We use $\textbf{1}$ and  $\textbf{0}$ to denote an all-one vector and an all-zero vector with a proper length, respectively. Sometimes, we use a subscript $n$ for $\textbf{1}$, i.e., $\textbf{1}_n$ to indicate its length $n$. The superscript of $\ba^t$ denotes the $t$-th iteration for $\ba$ \rv{in an iterative algorithm}. We use $[\ba]_n$ to denote the $n$-th element of vector $\ba$. \rv{The notation $D(\ba)$ represents a diagonal matrix with $\ba$ as its diagonal.}


\section{Approximate Message Passing with Unitary Transformation}


\rv{In this section, we briefly introduce the (UT)AMP algorithms, and show the close connection between UTAMP and AMP and their state evolution (SE).}

\subsection{(UT)AMP Algorithms}
The AMP algorithm \cite{Donoho2010a} was developed based on the loopy BP \cite{BPa}, \cite{Kschischang2001} for compressive sensing with model
\begin{equation}
\by=\bA\bx+\bw \label{ampeqa}
\end{equation}
where $\by$ is a measurement, $\bA$ is a known $M\times N$ measurement matrix, $\bw$ is a white Gaussian noise vector with distribution  $\N(\bw;\bm{0},\beta^{-1}\bI)$.
AMP enjoys low complexity and its performance can be rigorously characterized by a scalar state evolution in the case of large i.i.d. (sub)Gaussian $\bA$ \cite{javanmard2013state}.
However, for a generic $\bA$, the convergence of AMP cannot be guaranteed, e.g., AMP can easily diverge for non-zero mean, rank-deficient, correlated, or ill-conditioned matrix $\bA$ \cite{rangana2019convergence}, \cite{Guo2015UtAMP}.

Inspired by \cite{guo2013}, it was discovered in \cite{Guo2015UtAMP} that the AMP algorithm can still work well for difficult $\bA$. In \cite{Guo2015UtAMP}, instead of employing the original model (\ref{ampeqa}), AMP is applied to a unitary transform of (\ref{ampeqa}). As any matrix $\bA$ has an SVD  $\bA= \bU \bLambda \bV$, a unitary transformation with $\bU^H$ can be performed, yielding
\begin{equation}
\br=\bPhi \bx+\bm{\omega},
\label{r=uy}
\end{equation}
where $\br=\bU^H \by$, $\bPhi=\bU^H\bA=\mathbf{\Lambda}\bV$,
$\mathbf{\Lambda}$ is an $M\times N$ rectangular diagonal matrix, and $\bm{\omega} = \bU^H \bw$  is still a zero-mean Gaussian noise vector with the same covariance matrix  $\beta^{-1} \bI$ as $\bU^H$ is a unitary matrix. \rv{It is noted that in the case of a circulant matrix $\bA$, e.g., in frequency domain equalization, the matrix for unitary transformation can be simply the normalized discrete Fourier transform matrix, which allows more efficient implementation of the UTAMP algorithm \cite{guo2013}.} Then the vector stepsize AMP \cite{GAMP2011} shown in \textbf{Algorithm}~\ref{the vector stepsize AMP-table} can be applied to model (\ref{r=uy}), leading to the first version of the UTAMP algorithm, as shown in \textbf{Algorithm} \ref{UTAMPv1}. It is interesting that, with such a simple pre-processing, the robustness of AMP is remarkably enhanced, enabling it to handle difficult matrix $\bA$. \rv{ It is obvious that UTAMP can also be derived based on the loopy BP with the unitary transformed model \eqref{r=uy}.}

As discussed in \cite{Guo2015UtAMP}, applying an average operation to the two vectors $\bm{\tau}_x$ in Line 7 and $|\mathbf{\Phi}^H |^2   \bm{\tau}_s$ in Line 5 in \textbf{Algorithm} \ref{UTAMPv1} leads to the second version of UTAMP shown in \textbf{Algorithm} \ref{UTAMP}. Specifically, due to the average operation in Line 7 of \textbf{Algorithm} \ref{UTAMPv1}, $\bm{\tau}_x^t$ in Line 1  turns into a scaled all-one vector ${\tau}_x^t\bm{1}$. With $\bPhi=\mathbf{\Lambda}\bV$ \rv{and noting that $\bV$ is a unitary matrix}, it is not hard to show that
\begin{eqnarray}
\bm{\tau}_p  &=&  |\mathbf{\Phi} |^2  ({\tau}^t_x\bm{1})  \nonumber \\
&=&  \tau^t_x  \bm{\lambda},
\end{eqnarray}
which is Line 1 of \textbf{Algorithm} \ref{UTAMP}. Performing the average operation to vector $|\mathbf{\Phi}^H |^2   \bm{\tau}_s$, i.e.,
\begin{eqnarray}
<|\mathbf{\Phi}^H |^2 \bm{\tau}_s>=\frac{1}{N} \bm{\lambda}^T  \bm{\tau}_s
\end{eqnarray}
leads to Line 5 of \textbf{Algorithm} \ref{UTAMP}. \rv{It is worth highlighting} that the two average operations result in a significant reduction in computational complexity. Compared to \textbf{Algorithm} 1 and \textbf{Algorithm} 2, Line 1 and Line 5 of \textbf{Algorithm} \ref{UTAMP} do not \rv{involve} matrix-vector product operations, i.e., the number of matrix-vector products is reduced from 4 to 2 per iteration, which is a significant reduction as the complexity of AMP-like algorithms is dominated by matrix-vector products. \rv{Interestingly, the average operations also further enhance the stability of the algorithm from our finding. UTAMP version 2 converges for any matrix $\bA$ in the case of Gaussian priors \cite{Guo2015UtAMP}.} In many cases, the noise precision $\beta$ is unknown. The noise precision estimation can be incorporated into the UTAMP algorithms as in \cite{UTAMPSBL2019}. \rv{It is also worth mentioning that, VAMP involves the calculations of two "extrinsic" precisions (refer to Line 5 and Line 9 of Algorithm 1 in \cite{Sarkar2019}), which can be negative. To solve this problem, heuristic remedies can be used, e.g., taking the absolute value of the calculated precisions. In contrast, there is no such problem in the (UT)AMP algorithms.}

In the above \rv{(UT)AMP} algorithms, the function $g_x(\bq, \bm{\tau}_q )$ returns a column vector whose $n$-th
element, denoted as $[ g_x(\bq, \bm{\tau}_q ) ]_n$, is given by

\begin{equation}
[g_x(\mathbf{q}, \bm{\tau}_q ) ]_n
=
\frac{\int x_n p(x_n) \mathcal{N} (x_n ; q_n, \tau_{q_n})  d x_n }{\int  p(x_n) \mathcal{N} (x_n ; q_n, \tau_{q_n})  d x_n },
\label{g_x}
\end{equation}
where $p(x_n)$ is the prior of $x_n$. Equation (\ref{g_x}) can be interpreted as the minimum mean square error (MMSE) estimation of $x_n$ based on the following model
\begin{equation}
q_n = x_n + \varpi
\label{q_n}
\end{equation}
where $\varpi $ is a Gaussian noise with mean zero and variance $\tau_{q_n}$.
The function $g_x'(\bq,\bm{\tau}_q)$ returns a column vector and the $n$-th element is denoted by $[ g_x'(\bq, \bm{\tau}_q ) ]_n$, where the derivative is taken with respect to $q_n$. Note that $g_x(\bq,\bm{\tau}_q)$ can also be changed for MAP (maximum a posterior) estimation of $\bx$.

\subsection{State Evolution of (UT)AMP}

The performance of UTAMP can be characterized by the following simple recursion (for a more general matrix $\bA$ compared to AMP)
\begin{eqnarray}
	\tau^t &=& \frac{N}{\bm{1}^T \big(\bm{\lambda}./(v_x^{t}\bm{\lambda}+\beta^{-1}\bm{1})\big)} \label{UTAMPSE}   \\
	v_x^{t+1}&=&\mathbb{E}\left[\big|g_x(x+\sqrt{\tau^t} z,\tau^t) - x\big|^2 \right]
\end{eqnarray}
where $\beta^{-1}$ is the noise variance, $z$ is Gaussian with distribution $\N(z; 0, 1)$ and $x$ has a prior $p(x)$.

It is noted that, in the case of large i.i.d. Gaussian matrix $\bA$ with elements independently drawn from $\N(0, 1/ M)$, $\bm{\lambda}$ approaches a length-$M$ vector given by $\frac{N}{M} \bm{1}_M$ (assuming $M<N$). The SE of UTAMP is reduced to that of the AMP exactly, as in this case (\ref{UTAMPSE}) is reduced to
\begin{equation}
\tau^t=\frac{N}{M} v_x^t+\beta^{-1}.
\end{equation}

\begin{figure}[!t]
	\centering
	\includegraphics[width=0.45\textwidth] {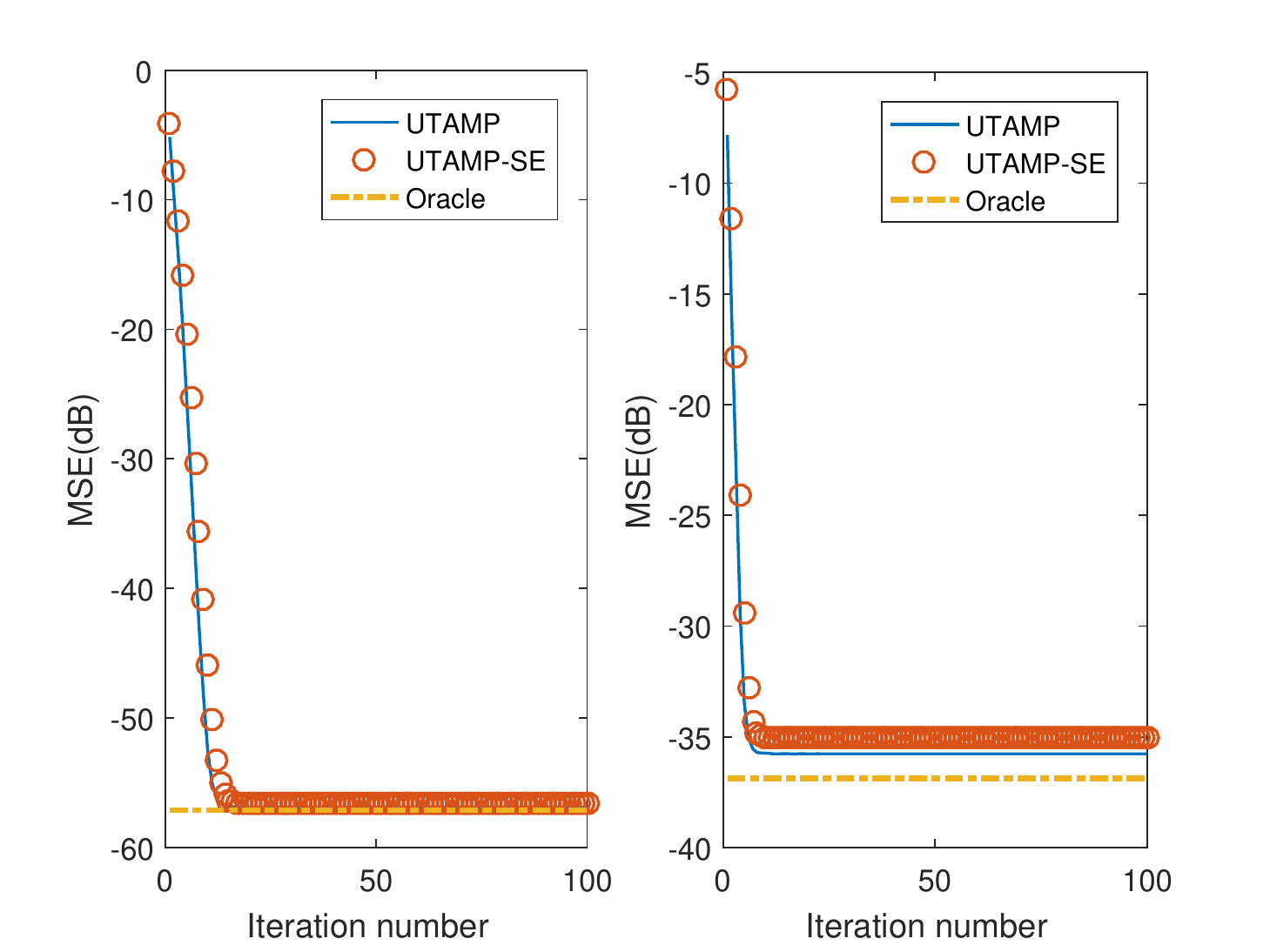}
	\centering
	\caption{Performance of UTAMP and its SE with a Bernoulli Gaussian prior for low-rank matrices (left) and non-zero mean matrices (right).}
	\label{fig:UTAMPSE}
\end{figure}

To demonstrate the SE of UTAMP, we assume that the measurement matrix has a size of $M = 800$ and $N = 1000$, the prior of the elements of $\bx$ is Bernoulli Gaussian $p(x)=0.9 \delta(x)+0.1\N(x;0,1)$, and the signal to noise ratio (SNR) is 50 dB. We generate non-zero mean matrices $\bA$ with elements independently drawn from $\N(10,1)$, and low rank matrices $\bA=\bB \bC$, where the size of $\bB$ and $\bC$ are $800\times 500$ and $500\times 1000$, respectively. Both $\bB$ and $\bC$ are i.i.d. Gaussian matrices with zero mean and unit variance. The mean squared error (MSE) of UTAMP and its SE are shown in Fig. \ref{fig:UTAMPSE} (the support-oracle MSE bound is also included \rv{for reference}), where we can see that the SE matches well the simulation performance. It is worth mentioning an interesting finding. In some cases, \rv{(UT)AMP} algorithms with the Bernoulli Gaussian prior cannot approach the support-oracle bound (e.g., the low-rank case), but UTAMP-SBL can still approach the bound as shown in \cite{UTAMPSBL2019}.


\section{Bilinear UTAMP}

In this section, the problem formulation for bilinear recovery is discussed, and the UTAMP based approximate inference algorithm Bi-UTAMP for bilinear recovery is derived. We start with the case of SMV, and then extend it to the case of MMV. The complexity of the algorithm is also analyzed.

\rv{
\subsection{Problem Formulation}
Different from \cite{Zhu2011CSMU}, we consider a Bayesian treatment of the bilinear recovery problem
\begin{eqnarray}
\by=\sum_{k=1}^K b_k\bA_k\bc+\bw, \label{eqnewa}
\end{eqnarray}
where $\bb\triangleq[b_1,...,b_K]^T$, $\bc$ and $\beta$ (the precision of the noise) are random variables with priors $p(\bb)$, $p(\bc)$ and $p(\beta)$, respectively. It is noted that, in the case of no a priori information available, $p(\bb)$, $p(\bc)$ and $p(\beta)$ can be simply chosen as non-informative priors. This also differs from the development of BAd-VAMP in \cite{Sarkar2019}, where both $\bb$ and $\beta$ are treated as unknown deterministic variables, and their values are estimated following the framework of expectation maximization (EM).
However, a Bayesian treatment of $\bb$ is more advantageous. In the case of a priori information available for $\bb$, a Bayesian method enables the use of the a priori information, which may be very helpful to improve the recovery performance. If no a priori information is known, a non-informative prior can be simply used. Moreover, in the context of iterative inference considered in this paper, the Bayesian treatment of $\bb$ is also different from that of the EM method in that only a point estimate of $\bb$ is involved in the iteration of the EM method, while a distribution of $\bb$ is involved in the iterative process of the method with Bayesian treatment\footnote{\rv{Even in the case of non-informative priors for the method with Bayesian treatment, they are still different in this way normally.}}.
Here, for simplicity, we take the SMV problem as example, but the extension of our discussion to the case of MMV is straightforward.
}

\rv{
The joint conditional distribution of $\bb$, $\bc$ and $\beta$ can be expressed as
\begin{equation}
p(\bb,\bc, \beta | \by) \propto p(\by|\bb,\bc,\beta)p(\bb)p(\bc)p(\beta).
\label{eq:eqnewb}
\end{equation}
We aim to find the a posterior distributions $p(\bb|\by)$ and $p(\bc|\by)$, and therefore their a posterior means that can be used as their estimates, i.e., $\hat \bb=\mathbb{E}(\bb|\by)$ and $\hat \bc=\mathbb{E}(\bc|\by)$. However, this is often intractable because high dimensional integration is required to compute the a posteriori distributions $p(\bb|\by)$ and $p(\bc|\by)$. As a result, we resort to the approximate Bayesian inference techniques.
}
\rv{
\subsection{Problem and Model Reformulation for Efficient UTAMP-Based Approximate Inference}
}

Similar to the lifting approach, we define $\bA\triangleq\left[\bA_{1},...,\bA_{K}\right]_{M\times NK}$, then the original bilinear model can be reformulated as
\begin{eqnarray}
\by=\bA\bx+\bw \label{eq:model2}
\end{eqnarray}
with \rv{the auxiliary variable}
\begin{eqnarray}
\bx = \bb\otimes \bc=
\begin{pmatrix}
b_1\bc\\
\vdots\\
b_K\bc
\end{pmatrix}
_{NK\times 1},\label{eq:x}
\end{eqnarray}
where $\bx$ can be indexed as
\begin{eqnarray}
\bx=\left[x_{1,1},...x_{N,1},...,x_{n,k},...x_{N,K}\right]^T\label{eq:newx}
\end{eqnarray}
with
\begin{eqnarray}
x_{n,k}=c_n b_k.
\end{eqnarray}

\rv{With an} SVD for matrix $\bA$, i.e.,
$\bA= \bU\bLambda\bV$,
performing unitary transformation \footnote{\rv{It is noted that performing the unitary transformation here is purely to facilitate the use of UTAMP. As $\bU^H$ is a unitary matrix, the transformation will not result in any loss. So the resultant algorithms will work with the transformed observation $\br$, instead of $\by$.}} yields $\br=\bPhi\bx+\bm{\omega}$,
where
$\br=\bU^{H}\by$, $\bPhi=\bLambda\bV$ has a size of $M \times NK$, and
$\bm{\omega}=\bU^{H}\bw$ is still white and Gaussian with the same precision $\beta$. \rv{Then define a new auxiliary variable $\bz=\bPhi\bx$ as in \cite{StretchZhang}, \cite{MengZhu}, \cite{XiangMing} and \cite{YuanZhang}. Later, we will see that the introduction of the auxiliary variables $\bx$ and $\bz$ facilitates the integration of UTAMP into the approximate Bayesian inference algorithm, which is crucial to achieving efficient and robust inference.}

\begin{table}[!hbp]
	\centering
	\caption{Distributions and factors in \eqref{eq:factor1}}
	\begin{tabular}{lll}
		\hline
		Factor & Distribution & Function\\
		\hline
		$f_{\br}$ & $p\left(\br|\bz,\beta\right)$ & $\N\left(\bz;\br,\beta^{-1}I\right)$\\
		$f_{\bz}$ & $p\left(\bz|\bx\right)$ & $\delta\left(\bz-\bPhi \bx\right)$ \\
		$f_{\bx}$ & $p(\bx|\bc,\bb)$ & $\delta\left(\bx-\bb\otimes\bc\right)$\\
		$f_{x_{n,k}}$ & $p\left(x_{n,k}|b_k,c_{n}\right)$ & $\delta\left(x_{n,k}-b_k c_{n}\right)$ \\
		$f_{\bc}$ & $p(\bc)$ & prior of $\bc$, e.g., prior promoting sparsity \\ 
		$f_{\bb}$ & $p(\bb)$ & prior of $\bb$\\ 
		$f_{\beta}$ & $p(\beta)$ & $\propto\beta^{-1}$ \\
		\hline
	\end{tabular}
\end{table}

\begin{figure}[!t]
\centering
\includegraphics[width=0.5\textwidth]{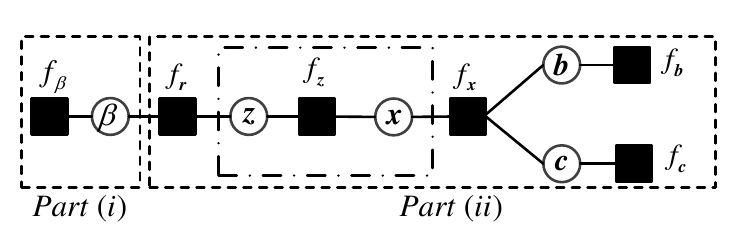}
\caption{\rv{Factor graph representation of \eqref{eq:factor1}.}}
\label{fig:Model}
\end{figure}

\rv{With the two latent variables $\bx$ and $\bz$, we have the following joint conditional distribution of $\bc,\bb,\bx,\bz, \beta$} and its factorization
\begin{eqnarray}
&&\!\!\!\!\!\!p(\bc,\bb,\bx, \bz, \beta | \br) \nonumber \\
&&\propto p(\br|\bz,\beta)p(\bz|\bx)p(\bx|\bb,\bc)p(\bc)p(\bb)p(\beta)\nonumber\\
&&\triangleq f_{\br}(\bz,\beta)f_{\bz}(\bz,\bx){f_{\bx}(\bx,\bb,\bc)} f_{\bc}(\bc)f_{\bb}(\bb) f_{\beta}(\beta).
\label{eq:factor1}
\end{eqnarray}
\rv{Hence our aim is to find the a posteriori distributions $p(\bc|\br)$ and $p(\bb|\br)$ and their estimates in terms of the a posteriori means, i.e., $\hat \bc=\mathbb{E}(\bc|\br)$ and $\hat \bb=\mathbb{E}(\bb|\br)$. It seems that, due to the involvement of two extra latent variables $\bx$ and $\bz$, the use of \eqref{eq:factor1} could be more complicated than that of \eqref{eq:eqnewb}, but it enables efficient approximate inference by incorporating UTAMP, as detailed later.} The probability functions and the corresponding factors (to facilitate the factor graph representation) are listed in Table 1, and a factor graph representation of \eqref{eq:factor1} is depicted in Fig. \ref{fig:Model}.

\rv{We follow the framework of structured variational inference (SVI) \cite{VMP1}, which can be formulated nicely as message passing with graphical models \cite{VMPnew}, \cite{Geiger05l}, \cite{Xing03ageneralized}, \cite{VMPJustin}. The trial function for the joint conditional distribution function $p(\bc,\bb,\bx, \bz, \beta | \br)$ in \eqref{eq:factor1} is chosen as
\begin{equation}
\tilde{q}(\bb, \bc, \bx, \bz, \beta)= \tilde{q}(\beta) \tilde{q}(\bb, \bc, \bx, \bz).
\end{equation}
The employment of this trial function corresponds to a partition of the factor graph in Fig. \ref{fig:Model} \cite{VMPJustin}, i.e., $\tilde{q}(\beta)$ and $\tilde{q}(\bb, \bc, \bx, \bz)$ are associated respectively with the subgraphs denoted by Part (i) and Part (ii), where the variable node $\beta$ is external to Part (ii). With SVI, the variational lower bound
\begin{eqnarray}
\mathcal L \big(\tilde{q}(\bb, \bc, \bx, \bz, \beta) \big) =&& \nonumber \\ &&\!\!\!\!\!\!\!\!\!\!\!\!\!\!\!\!\!\!\!\!\!\!\!\!\!\!\!\!\!\!\!\!\!\!\!\!\!\!\!\!\!\!\!\!\!\!\!\!\!\!\!\!\mathbb{E}\big[\mathrm{log}(p(\bc,\bb,\bx, \bz, \beta, \br))] - \mathbb{E}\big[\mathrm{log}(\tilde{q}(\bb, \bc, \bx, \bz, \beta))]
\end{eqnarray}
is maximized with respect to the trial function, so that the following Kullback-Leibler divergence
\begin{eqnarray}
\mathcal{KL}\big(\tilde{q}(\beta) \tilde{q}(\bb, \bc, \bx, \bz)|| p(\bb,\bc,\bx, \bz, \beta | \br)\big),
\end{eqnarray}
is minimized, which leads to the approximation (by integrating out $\beta$)
\begin{equation}
\tilde{q}(\bb, \bc, \bx, \bz) \approx p(\bb,\bc,\bx, \bz | \br).
\end{equation}
From the above, by integrating out $\bc, \bx$ and $\bz$, it is expected that the marginal $\tilde{q}(\bb) \approx p(\bb | \br) $, and similarly, by integrating out $\bb, \bx$ and $\bz$, $\tilde{q}(\bc) \approx p(\bc | \br) $. In terms of structured variation message passing \cite{VMPJustin}, the computation of $\tilde{q}(\bb, \bc, \bx, \bz)$ corresponds to BP in the subgraph shown in Part (ii) of the factor graph in Fig. \ref{fig:Model}, except the function node $f_{\br}$ because it connects an external variable node $\beta$ \cite{VMPJustin}. It is noted that the BP message passing between $\bz, f_z$ and $\bx$ (i.e., BP in the dash-dotted box in Fig. \ref{fig:Model}) can be difficult and computational intensive. Fortunately, AMP, derived based on loopy BP (which in this case is actually UTAMP as the unitary transformation has already been performed previously) is an excellent replacement to accomplish the BP message passing for the dash-dotted box efficiently. In addition, we may have difficulties with the priors $p(\bb)$ and $p(\bc)$ (corresponding to the factors $f_{\bb}$ and $f_{\bc}$ in Fig. \ref{fig:Model}) as they may not be friendly, resulting in intractable BP messages. This can be handled with EP, which has been widely used in the literature to solve similar problems. At the variable node $\bc$ (or $\bb$), we can obtain an approximate marginal about $\bc$ (or $\bb$) through an iterative process with moment matching \cite{EP}, thereby an approximation to the a posteriori mean $\mathbb{E}(\bc | \br)$ (or $\mathbb{E}(\bb | \br)$), which can be served as our estimate.
}

\rv{It is noted that, all inference methods mentioned above including VI, EP, and UTAMP involve an iterative process (but with a different hierarchy), and the multiple iterative processes can be simply combined as a single one. In terms of message passing, this is to carry out a forward message passing process and a backward message passing process in Fig. \ref{fig:Model} as an iteration. Thanks to the incorporation of UTAMP to handle the BP in the dashed-dotted box in Fig. \ref{fig:Model}, this leads to an efficient and robust approximate inference algorithm with details elaborated in next section.   	
}

\rv{
\subsection{Derivation of the Message Passing Algorithm}
}

\rv{
In this section, we \rev{detail} the forward and backward message passing in Fig. \ref{fig:Model} according to the principle of structured variational message passing \cite{VMP1}, \cite{VMPnew}, \cite{VMPJustin} and EP. Throughout this paper, we use the notation $m_{n_a \rightarrow  n_b} (h)$ to denote a message passed from node $n_a$ to node $n_b$, which is a function of $h$.
}

\rv{
\subsubsection{Message Computations at Nodes $\textbf{x}$, $f_\textbf{z}$, $\textbf{z}$ and $f_\textbf{r}$} Treat $\bx$, $f_{\bz}$ and $\bz$ as a module, shown by the dash-dotted box in Fig. \ref{fig:Model}. In the backward direction, with the incoming messages from the factor nodes $f_{\bx}$ as the input, the module needs to output the message $m_{\bz  \rightarrow  f_{\br}} (\bz)$. In the forward direction, with the incoming messages from the factor node $f_{\br}$ as input, the module needs to output the message $m_{\bx  \rightarrow  f_{\bx}} (\bx)$. This is the most computational intensive part of the approximate inference method, and it can be efficiently handled with UTAMP as mentioned earlier. Considering the structure of $\bx$ shown in \eqref{eq:x},} we divide the length-$NK$ vector $\bx$ into $K$ length-$N$ vectors $\{\bx_k, k=1, ..., K\}$, i.e.,
\begin{equation}
\bx=\left[\bx_1^T,..., \bx_K^T\right]^T.
\end{equation}
Due to this, the UTAMP algorithms in Section II cannot be applied directly, but the derivation still follows that of the UTAMP algorithms exactly.

Note that the size of matrix $\bPhi$ is $M \times NK$. We partition it into $K$ sub-matrices $\{\bPhi_k, k=1, ..., K\}$, each with a size of $M \times N$, i.e.,
\begin{equation}
\bPhi=\left[\bPhi_1,..., \bPhi_K \right].
\end{equation}
Then we define $K$ vectors $\{\bphi_k, k=1, ..., K\}$, each with a length of $M$, i.e.,
\begin{equation}
\bphi_k=|\bPhi_k|^2 \textbf{1}_N.
\end{equation}
With the above definitions, we have the following model
\begin{equation}
\br=\sum_{k=1}^K\bPhi_k\bx_k+\bm{\omega}.
\end{equation}

\rv{We first investigate the backward message passing}.  Assume that the incoming message from factor node $f_x$ is available, which is the mean and variance of $\bx_k$. Following UTAMP, we assume that the elements of $\bx_k$ have a common variance $v_{\bx_k}$, and the computation of $v_{\bx_k}$ will be detailed later. The mean of $\bx$ is denoted by $\hat \bx$.
Then we calculate two vectors $\bnu_{\bp}$ and $\bp$ as
\begin{eqnarray}
&&\bnu_{\bp}= \sum_{k=1}^K \bphi_k v_{\bx_k} \label{eq:np}\\
&& \bp=\sum_{k=1}^K\bPhi_k\hat\bx_k-\bnu_{\bp}\cdot\bs,\label{eq:hatp}
\end{eqnarray}
where $\bs$ is a vector, which is computed in the last iteration. According to the BP derivation of (UT)AMP \footnote {\rv{UTAMP also allows a loopy BP derivation that is the same as AMP, except that the derivation is based on the unitary transformed model.}},
\begin{equation}
m_{\bz  \rightarrow  f_{\br}} (\bz)= m_{f_{\bz} \rightarrow  \bz} (\bz)= \N\left(\bz;\bp, D(\bnu_p)\right).
\end{equation}

\rv{It is noted that the factor node $f_{\br}$ connects the external variable node $\beta$. According to the rules of the structured variational message passing \cite{VMPJustin}, the message $m_{f_{\br} \rightarrow \beta}(\beta)$  can be computed as
\begin{equation}
m_{f_{\br} \rightarrow \beta}(\beta)  \propto \exp \left\{ \int_{\bz} \mathfrak{b}(\bz) {\log  f_{\br} } \right\},
\end{equation}
where  $\mathfrak{b}(\bz)$ is the the approximate marginal of $\bz$, i.e.,
\begin{equation}
\begin{aligned}
\mathfrak{b}(\bz) &\propto m_{f_{\br} \rightarrow \bz}(\bz) m_{\bz \rightarrow f_{\br}}(\bz)\\
&=\N(\bz; \hat \bz, D(\bnu_{\bz}))
\end{aligned}
\end{equation}
with}

\begin{eqnarray}
&&\bnu_{\bz}=\bm{1}./\left(\bm{1}./\bnu_{\bp}+\hat\beta\textbf{1}_{M}\right) \label{eq:np1}  \\ 
&&\hat \bz=\bnu_{\bz}\cdot\left( \bp./\bnu_{\bp}+\hat\beta\br\right) \label{eq:np1a}
\end{eqnarray}
where $\hat\beta$ is the approximate a posteriori mean of the noise precision $\beta$ in the last iteration. Note that there may be zero elements in $\bnu_{\bp}$. To avoid the potential numerical problem, the above equations can be rewritten as
\begin{eqnarray}
&&\bnu_{\bz} = \bnu_p./(\bm{1}+\hat\beta \bnu_p)  \\
&&\hat \bz=(\hat\beta\bnu_p\cdot \bm{r}+\bm{p})./(\bm{1}+\hat\beta\bnu_p).
\end{eqnarray}
\rv{It is noted that in the above derivation, the message $m_{f_{\br} \rightarrow \bz}(\bz)$ is required, which turns out to be Gaussian, i.e., $m_{f_{\br} \rightarrow z}(\bz)=\N(\bz,\br, \hat\beta^{-1})$, and its derivation is delayed to \eqref{eq：eqnewc}.  Then, it is not hard to show that the message
\begin{equation}
m_{f_{\br}\rightarrow \beta} (\beta) \propto  {\beta}^{M}\exp \{ -{\beta} (|| \br - \hat \bz ||^2 + \textbf{1}^T \bnu_{\bz})  \} .
\end{equation}
This is the end of the backward message passing.}

\rv{Next, we investigate the forward message passing. According to the rules of the structured variational message passing and noting that $f_{\br}$ connects the external variable node $\beta$, we have
\begin{eqnarray}
m_{f_{\br} \rightarrow \bz}(\bz)  &\propto& \exp \left\{ \int_{\beta} \mathfrak{b}(\beta) {\log  f_{\br} } \right\}, \nonumber \\
&\propto& \N(\bz; \br, \hat\beta^{-1}) \label{eq：eqnewc}
\end{eqnarray}
with
\begin{eqnarray}
 \mathfrak{b}(\beta) &\propto& m_{f_{\br} \rightarrow \beta}(\beta) f_{\beta} \\ \nonumber
 &\propto&  {\beta}^{M-1}\exp \{ -{\beta} \big(|| \br - \hat \bz ||^2 + \textbf{1}^T \bnu_{\bz} \big) \},
\end{eqnarray}
and	
}
\begin{eqnarray}
\hat\beta= \rv{ \int_{\beta} \beta \mathfrak{b}(\beta) =}   \frac{M}{\left\|\br-\hat \bz\right\|^2+\textbf{1}^T\bnu_{\bz}}, \label{eq:lambda}
\end{eqnarray}
where we slightly abuse the use of the notation $\hat\beta$ as we do not distinguish it from the last iteration. \rv{The result for $\hat\beta$  coincides with the result in \cite{VAMPE1} and  \cite{VAMPE}.}


\rv{The message $m_{f_{\br} \rightarrow \bz}(\bz)$ is input to the dash-dotted box in Fig. \ref{fig:Model}. The Gaussian form of the message suggests the following model
\begin{equation}
\br=\bz+ \bw',
\end{equation}
where the noise $\bw'$ is Gaussian with mean zero and precision $\hat\beta$. This allows seamless connection with the forward recursion of UTAMP.
}	
According to UTAMP, we update the intermediate vectors  $\bnu_{\bs}$ and $ \bs$ by
\begin{eqnarray}
&&\bnu_{\bs}=\bm{1}./(\bnu_{\bp}+\hat\beta^{-1}\textbf{1})\\
&& \bs=\bnu_{\bs}\cdot\left(\br-\bp\right)\label{eq:s}.
\end{eqnarray}
Then calculate vectors $\nu_{\bq_k}$ and $\hat \bq_k$ for $k=0, ..., K$ with
\begin{eqnarray}
\nu_{\bq_k}&=&1/\left<|\bPhi_k^H|^2\bnu_{\bs}\right>\label{eq:nuq}\\
\bq_k&=&\hat \bx_k+\nu_{\bq_k}\bPhi_k^H \bs.
\end{eqnarray}
The messages $\bq_k$ and $\nu_{\bq_k}$ are the mean and variance of $\bx_k$.
According to the BP derivation of (UT)AMP,
\begin{equation}
m_{\bx \rightarrow f_{\bx}}(\bx)= \N(\bx; \bq, D(\bnu_{\bq}))
\end{equation}
with
\begin{eqnarray}
\bq&=&[\bq_1^T, ..., \bq_K^T]^T  \\
\bnu_{\bq}&=&[\nu_{\bq_1}, ..., \nu_{\bq_K}]^T \otimes \textbf{1}_N,
\end{eqnarray}
which is the output of the dash-dotted box in Fig. \ref{fig:Model}. This is the end of the forward message passing.

\subsection{Message Computations at Nodes $f_\textbf{x}$, $\textbf{b}$ and $\textbf{c}$}

\begin{figure}[!t]
	\centering
	\includegraphics[width=0.25\textwidth]{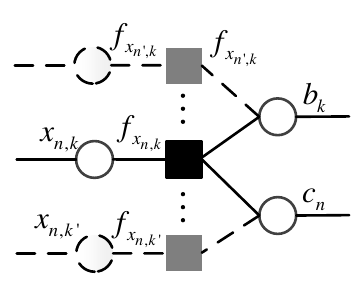}
	\centering
	\caption{\rv{Factor graph representation for  $f_{x_{n,k}}(c_n, b_k)$.}}
	\label{fig:ModelScalar}
\end{figure}

We note that the function $f_{\bx}(\bx,\bc,\bb)$ can be further factorized, i.e.,
\begin{equation}
f_{\bx}(\bx,\bc,\bb)=\prod\nolimits_{n,k}f_{x_{n,k}}(b_k,c_n),
\end{equation}
and the factor $f_{x_{n,k}}(c_n, b_k)$ is shown in Fig. \ref{fig:ModelScalar} with solid lines, which will be used to derive the forward and backward message computations.

\rv{We first investigate the forward message passing, where the message $m_{\bx \rightarrow f_{\bx}}(\bx)$ is available from the dash-dotted box .} The $n$th entry of $\bq_k$ is denoted by $q_{n,k}$, \rv{then we have} $m_{x_{n,k}\to f_{x_{n,k}}}(x_{n,k})=\N(x_{n,k}; q_{n,k},\nu_{\bq_{k}})$ and the factor $f_{x_{n,k}}=\delta\left(x_{n,k}-b_k c_{n}\right)$.

\rv{To compute the message $m_{f_{x_{n,k}}\to c_n}(c_n)$ with BP at factor node $f_{x_{n,k}}$, we need to integrate out $x_{n,k}$ and $b_k$. However, due to the multiplication of $b_k$ and $c_n$, the message will be intractable even if the incoming message $m_{b_k \to f_{x_{n,k}}} (b_k)$ is Gaussian.}
To solve this, we first apply BP and eliminate the variable $x_{n,k}$ to get an intermediate function node $\tilde f_{x_{n,k}}(c_{n},b_k)$, i.e.,
\begin{eqnarray}
\tilde f_{x_{n,k}}(c_{n},b_k)&=&\int_{x_{n,k}} m_{x_{n,k}\to f_{x_{n,k}}}(x_{n,k}) \cdot f_{x_{n,k}} \nonumber\\
&=&\N\left(c_{n}b_k; q_{n,k},\nu_{\bq_{k}}\right)\label{eq:fx_td}.
\end{eqnarray}
This turns the function node $f_{x_{n,k}}$ with the hard constraint $\delta\left(x_{n,k}-b_k c_{n}\right)$ to a 'soft' function node, enabling the use of \rv{variational inference} to handle $c_{n}$ and $b_k$.  With the intermediate local function $\tilde f_{x_{n,k}}(b_k,c_{n})$, we can calculate the outgoing message from $f_{x_{n,k}}$ to $c_n$ as
\begin{eqnarray}
m_{f_{x_{n,k}}\to c_n}(c_n)&=&\exp\left\{\int_{b_{k}} \mathfrak{b}(b_{k}) \log\tilde f_{x_{n,k}}\right\}\nonumber\\
&=&\N\left(c_n;\vec c_{n,k},\vec\nu_{c_{n,k}}\right)
\end{eqnarray}
where
\begin{eqnarray}
\vec c_{n,k}=\frac{q_{n,k}\hat b_{k}^*} {|\hat b_{k}|^2+\nu_{b_{k}}}, \\
\vec\nu_{c_{n,k}}=\frac{\nu_{\bq_{k}}} {|\hat b_{k}|^2+\nu_{b_{k}}},\label{eq:x2b}
\end{eqnarray}
with $\hat b_{k}$ and $\nu_{b_{k}}$ being the approximate a posteriori mean and variance of $b_k$, which are computed in \eqref{eq:bc1} and \eqref{eq:bc}. It is noted that, in the case of $b_1=1$, we simply set $\hat b_1=1$ and $\nu_{b_1}=0$. \rv{With BP and referring to Fig. \ref{fig:ModelScalar}}, the message $m_{c_n \to f_{\bc}}(c_n)$ can be represented as
\begin{eqnarray}
m_{c_n \to f_{\bc}}(c_n) =\N\left(c_n;\vec c_n,\vec\nu_{c_{n}}\right)
\end{eqnarray}
with
\begin{eqnarray}
\vec\nu_{c_{n}}=1/\sum_{k=1}^K \frac{1}{\vec\nu_{c_{n,k}}} \\
\vec{c}_{n}=\vec\nu_{c_{n}} \sum_{k=1}^K \frac{ {\vec c_{n,k}}}{\vec\nu_{c_{n,k}}}.
\end{eqnarray}
So, the marginal of $c_n$ $(n=1, ..., N)$ can be expressed as
\begin{eqnarray}
\mathfrak{b}(c_n)=\int_{\bc \vee {c_n}} \prod_n m_{c_n \to f_{\bc}}(c_n) f_{\bc}.
\end{eqnarray}
\rv{As mentioned earlier, according to EP, the marginal} is projected to be Gaussian through moment matching, i.e.,
\begin{eqnarray}
\mathfrak{b'}(c_n) =\N\left(c_n;\hat c_n,\nu_{c_{n}}\right)
\end{eqnarray}
with
\begin{eqnarray}
\hat c_n&=&\mathbb{E}\Big[c_n|\{\vec\nu_{c_n},\vec c_n\}, f_{\bc}\Big] \label{eq:c1a}\\
\nu_{c_n}&=&\mathbb{V}\text{ar}\Big[c_n|\{\vec\nu_{c_n},\vec c_n\}, f_{\bc}\Big]\label{eq:c1b},
\end{eqnarray}
which are a posterior mean and variance of $c_n$ based on the prior $f_{\bc}$ and the following pseudo observation model \rv{ \cite{MengZhu}, \cite{XiangMing}, \cite{Xiaodong}}
\begin{eqnarray}
\vec c_n= c_n + w'_n, 
\end{eqnarray}
with $w'_n$ denoting a Gaussian noise with mean 0 and variance $\vec \nu_{c_n}$.

Similarly, we can calculate the message from $f_{x_{n,k}}$ to $b_k$, i.e.,
\begin{eqnarray}
m_{f_{x_{n,k}}\to b_k}(b_k) = \N\left(b_k;\vec b_{n,k},\vec\nu_{b_{n,k}}\right)
\end{eqnarray}
where
\begin{eqnarray}
\vec b_{n,k}=\frac{q_{n,k}\hat c_{n}^*} {|\hat c_{n}|^2+\nu_{c_{n}}}, \\
\vec\nu_{b_{n,k}}=\frac{\nu_{\bq_{k}}} {|\hat c_{n}|^2+\nu_{c_{n}}}
\end{eqnarray}
with $\hat c_{n}$ and $\nu_{c_{n}}$ being the approximate a posteriori mean and variance of  $c_n$, which are updated in \eqref{eq:c1a} and \eqref{eq:c1b}.
Then with BP, the message $m_{b_k \to f_{\bb}}(b_k)$ can be expressed as
\begin{eqnarray}
m_{b_k \to f_{\bb}}(b_k)= \N\left(b_k;\vec b_k,\vec\nu_{b_k}\right)
\end{eqnarray}
with
\begin{eqnarray}
\vec\nu_{b_k}&=&1/\sum_{n=1}^N\frac{1}{\vec\nu_{b_{n,k}}}\\ 
\vec b_k &=&\vec\nu_{b_k} \sum_{n=1}^N \frac{\vec b_{n,k}}{\vec\nu_{b_{n,k}}}.
\end{eqnarray}
Then we can compute the marginal of each $b_k$,
\begin{eqnarray}
\mathfrak{b}(b_k)=\int_{\bb \vee {b_k}} \prod_k m_{b_k \to f_{\bb}}(b_k) f_{\bb}.
\end{eqnarray}
Similarly, it is then  projected to be Gaussian, i.e.,
\begin{eqnarray}
\mathfrak{b'}(b_k) =\N\left(b_k;\hat b_k,\nu_{b_{k}}\right)
\end{eqnarray}
with
\begin{eqnarray}
\hat b_k&=&\mathbb{E}\Big[b_k|\{\vec\nu_{b_k},\vec b_k\}, f_{\bb}\Big]\label{eq:bc1}\\
\nu_{b_k}&=&\mathbb{V}\text{ar}\Big[b_k|\{\vec\nu_{b_k},\vec b_k\}, f_{\bb}\Big]\label{eq:bc},
\end{eqnarray}
which are the a posteriori mean and variance of $b_k$ based on the prior $f_{\bb}$ and the following pseudo observation model
\begin{eqnarray}
\vec b_k= b_k + w''_k 
\end{eqnarray}
with $w''_k$ denoting a Gaussian noise with mean 0 and variance $\vec \nu_{b_k}$. It is noted that, in the case of $b_1=1$, we simply set $\hat b_1=1$ and $\nu_{b_1}=0$. \rv{This is the end of the forward message passing.}

\begin{algorithm}
	\caption{Bi-UTAMP for SMV}
	\textbf{Unitary transform}: $\br=\bU^H \by =\bPhi \bx+\bomega$, where $\bA_{M \times NK}=\bU \bLambda \bV$, $\bPhi=\bU^H \bA=\bLambda \bV$, and $\bx = \bb\otimes \bc$ with $\bb=[b_1, ..., b_K]^T$ and $\bc=[c_1, ..., c_N]^T$. \\
	Let $\bPhi=\left[\bPhi_1, ..., \bPhi_K \right]$, $\bphi_k=|\bPhi_k|^2 \textbf{1}_N$, and $\bx=\left[\bx_1^T, ..., \bx_K^T\right]^T$, $k= 1,...K$ and $n=1, ...,N$.   \\
	\textbf{Initialize} $\hat b_k$, $\nu_{b_k}=1$, $\nu_{\bx_k}=1$, $\hat\bx_k=\textbf{0}$, $\bs=\mathbf{0}$ and $\hat \beta=1$. \\
	\textbf{Repeat}
	\begin{algorithmic}[1]
		\STATE$\bnu_{\bp}= \sum_k \bphi_k \nu_{\bx_k}$\\
		\STATE$\bp=\sum_k \bPhi_k\hat\bx_k-\bnu_{\bp}\cdot\bs$\\
		\STATE$\bnu_{\bz} = \bnu_p./(\bm{1}+\hat\beta \bnu_p)$ \\
		\STATE$\hat \bz=(\hat\beta\bnu_p\cdot \bm{r}+\bm{p})./(\bm{1}+\hat\beta\bnu_p) $ 
		\STATE$\hat\beta={M}/{(\left\|\br-\hat \bz\right\|^2+\textbf{1}^T\bnu_{\bz})} $ \\
        \STATE$\bnu_{\bs}=\bm{1}./(\bnu_{\bp}+\hat\beta^{-1}\textbf{1}_{M})$ \\
		\STATE$\bs=\bnu_{\bs}\cdot(\br-\bp)$ \\
		\STATE$\forall k: \nu_{\bq_k}=1/\left<|\bPhi_k^H|^2\bnu_{\bs}\right>$ \\
        \STATE$\forall k: \bq_k=\hat \bx_k+\nu_{\bq_k}\bPhi_k^H \bs$ \\

        (In the case of $b_1=1$, set $\hat b_1=1$ and $\nu_{b_1}=0$.) \\
        \STATE$\forall k:\vec \bc_{k}={\bq_{k}\hat b_{k}^*} /({|\hat b_{k}|^2+\nu_{b_{k}}})$ \\
        \STATE$\forall k:\vec\bnu_{\bc_{k}}=\textbf{1}_N{\nu_{\bq_{k}}}/ ({|\hat b_{k}|^2+\nu_{b_{k}}})$ \\
        \STATE$ \vec\bnu_{c}=\textbf{1}_N./(\sum_k\textbf{1}_N./\vec\bnu_{c_{k}})$ \\
        \STATE$\vec{\bc}=\vec\bnu_{\bc}\cdot\sum_k(\vec \bc_{k}./\vec\bnu_{\bc_{k}})$ \\
        \STATE$\forall n:\hat c_n=\mathbb{E}[c_n|\vec\bnu_{\bc},\vec \bc, f_{\bc}]$ \\
        \STATE$\forall n:\nu_{c_n}=\mathbb{V}\text{ar}[c_n|\vec\bnu_{\bc},\vec \bc, f_{\bc}]$ \\

        \STATE$\bnu_{\bc}=<[\nu_{c_1}, ...,\nu_{c_N}]> \textbf{1}_N$, and $\hat \bc= [\hat c_1, ...,\hat c_N]^T$\\

        \STATE$\forall k:\vec\bnu_{\bb_{k}}={\nu_{\bq_{k}}}\textbf{1}_N./ ({\left|\hat \bc\right|^2+\bnu_{\bc}})$ \\
        \STATE$\forall k:\!\vec \bb_{k}={\bq_{k}\cdot\hat \bc^*}./({\left|\hat \bc\right|^2+\bnu_{\bc}})$\\ 

        \STATE$\forall k:\vec\nu_{b_{k}}=(\textbf{1}_N^T (\textbf{1}_N./\vec\bnu_{\bb_k}))^{-1}$ \\
        \STATE$\forall k:\vec{b}_{k}=\vec\nu_{b_{k}} \textbf{1}_N^T(\vec \bb_{k}./\vec\bnu_{b_{k}})$ \\
        \STATE$\forall k:\hat b_k=\mathbb{E}[b_k|\{\vec\nu_{b_k},\vec b_k\}, f_{\bb}] $ \\
        \STATE$\forall k:\nu_{b_k}=\mathbb{V}\text{ar}[b_k|\{\vec\nu_{b_k},\vec b_k\}, f_{\bb}]$ \\

        (In the case of $b_1=1$, set $\hat b_1=1$ and $\nu_{b_1}=0$.) \\

        \STATE$\forall k: \cev\bnu_{\bb_k}=\big(\nu_{b_k} \vec\bnu_{ \bb_k}\big)./\big( \vec\bnu_{\bb_k}-\nu_{b_k} \textbf{1}_N\big) $
        \STATE$\forall k:\cev\bb_k=\big(\hat b_k \vec\bnu_{\bb_k}- \nu_{b_k}\vec\bb_k  \big)./\big( \vec\bnu_{\bb_k}-\nu_{b_n} \textbf{1}_N\big)$\\

        \STATE$\forall k: \cev\bnu_{c_k}\!=\!\left(\textbf{1}./\bnu_c-\!\textbf{1}./\vec\bnu_{\bc_k}\right)^{.-1}$ \\
        \STATE$\forall k: \cev\bc_k=\cev\bnu_{c_k}\cdot\left(\hat\bc./\bnu_c-\!\vec\bc_k./\vec\bnu_{\bc_k}\right)$ \\
        \STATE$\forall k: \cev \bx_k=\cev\bb_k\cdot\cev\bc_k$\\ 
        \STATE$\forall k: \cev \bnu_{\bx_k}=|\cev\bb_k|^2\cdot\cev\bnu_{\bc_k}+\cev\bnu_{\bb_k}\cdot\left|\cev\bc_k\right|^2+
        \cev\bnu_{\bb_k}\cdot\cev\bnu_{\bc_k}$\\ 
        \STATE$\forall k:\bnu_{\bx_k}=\left(1/\nu_{\bq_k}\textbf{1}_N+\textbf{1}./\cev \bnu_{\bx_k}\right)^{.-1}$ \\
        \STATE$\forall k:\hat \bx_k=\bnu_{\bx_k} \cdot \left( 1/\nu_{\bq_k}\bq_k+\cev\bx_k./\cev\bnu_{\bx_k}\right)$ \\
        \STATE$\forall k: \nu_{\bx_k}=<\bnu_{\bx_k}>$
	\end{algorithmic}
	\textbf{Until terminated}
	\label{BiUTAMPSMV}
\end{algorithm}

\rv{Next, we investigate the backward message passing.} \rv{According to the rule of EP, the backward message
\begin{equation}
m_{b_k \to f_{x_{n,k}}}(b_k)= \frac{\mathfrak{b'}(b_k)}{m_{f_{x_{n,k} \to b_k}}(b_k)}.
\end{equation}
They are represented collectively as $m_{\bb\to f_{\bx}}(\bb)$}, which is Gaussian with mean $\cev{ \bb}$ and variance \rv{$D(\cev\bnu_{ \bb})$}. With the factor graph shown in Fig. \ref{fig:ModelScalar}, the mean and variance can be calculated as
\begin{eqnarray}
\cev\bnu_{ \bb}&=&\left(\left(\textbf{1}./\bnu_{ \bb}\right)\otimes \textbf{1}_{N}-\textbf{1}./\vec\bnu_{ \bb}\right)^{.-1} \nonumber\\
&=& \big((\bnu_{\bb} \otimes \textbf{1}_N) \cdot \vec\bnu_{ \bb}\big)./\big( \vec\bnu_{ \bb}-(\bnu_{\bb} \otimes \textbf{1}_N)\big) \label{eq:rb}\\
\cev{\bb}&=&\cev\bnu_{ \bb}\cdot\left(\big(\hat{\bb}./\bnu_{ \bb}\big)\otimes \textbf{1}_{N}-\vec{\bb}./\vec\bnu_{\bb}\right),\nonumber \\
&=& \big((\hat\bb \otimes \textbf{1}_N) \cdot \vec\bnu_{ \bb}-\vec\bb\cdot(\bnu_{\bb} \otimes \textbf{1}_N) \big)./\big( \vec\bnu_{ \bb}-(\bnu_{\bb} \otimes \textbf{1}_N)\big)\nonumber\\ \label{eq:rb}
\end{eqnarray}
where $\bnu_{\bb}=[\nu_{b_1},...,\nu_{b_K}]^T$, $\hat{\bb}=[\hat b_1, ..., \hat b_K]^T$, $\big[\vec \bnu_{\bb}\big]_{(k-1)N+n}=\vec \nu_{b_{n,k}}$ and $[~\vec {\bb}~]_{(k-1)N+n}=\vec {b}_{n,k}$.

Similarly, the message $m_{\bc\to f_{\bx}}(\bc)$ is also Gaussian with mean $\cev\bc$, and variance \rv{$D(\cev\bnu_{c})$}, which can be calculated as
\begin{eqnarray}
&&\cev\bnu_{c}=\left(\textbf{1}_{K}\otimes\left(\textbf{1}./\bnu_c\right)-\textbf{1}./\vec\bnu_{\bc}\right)^{.-1} \\
&&\cev\bc=\cev\bnu_{c}\cdot\left(\textbf{1}_{K}\otimes\left(\hat\bc./\bnu_c\right)-\vec\bc./\vec\bnu_{\bc}\right),
\end{eqnarray}
where $\bnu_{\bc}=[\nu_{c_1},...,\nu_{c_N}]^T$, $\hat\bc=[\hat c_1, ..., \hat c_N]^T$, $\big[\vec \bnu_{\bc}\big]_{(n-1)K+k}=\vec \nu_{c_{n,k}}$ and $\big[\vec {\bc}\big]_{(n-1)K+k}=\vec {c}_{n,k}$.
Then, the backward message $m_{f_{\bx}\to \bx}(\bx)=\N\left(\bx;\cev \bx,\cev \bnu_{\bx}\right)$ with
\begin{eqnarray}
&&\cev \bx=\cev\bb\cdot\cev\bc \label{eq:theta}\\
&&\cev \bnu_{\bx}=|\cev\bb|^2\cdot\cev\bnu_{\bc}+\cev\bnu_{\bb}\cdot\left|\cev\bc\right|^2+
\cev\bnu_{\bb}\cdot\cev\bnu_{\bc} \label{eq:nuTheta},
\end{eqnarray}
where $\cev\bx=[\cev\bx_1^T,..., \cev\bx_K^T]^T$ and $\cev\bnu_{\bx}=[\cev\bnu_{\bx_1}^T,..., \cev\bnu_{\bx_K}^T]^T$.
The backward message is combined with \rv{the message $m_{\bx \to f_{\bx}} (\bx)$ (the output of the dash-dotted box in last iteration)} i.e.,
\begin{eqnarray}
\bnu_{\bx_k}&=&\left(1/\nu_{\bq_k}\textbf{1}_N+\textbf{1}./\cev \bnu_{\bx_k}\right)^{.-1} \\
\hat \bx_k&=&\bnu_{x_k}\cdot \left( 1/\nu_{\bq_k}\bq_k+\cev\bx_k./\cev\bnu_{\bx_k}\right) \\
\nu_{\bx_k}&=&<\bnu_{\bx_k}>
\end{eqnarray}
which are then passed to \rv{the dash-dotted box as input. This is the end of the backward message passing.}

The \rv{approximate inference} algorithm is called Bi-UTAMP for SMV, and it can be organized in a more succinct form,  which is summarized in \textbf{Algorithm} \ref{BiUTAMPSMV}.

\subsection{Extension to MMV}

In this section, we extend Bi-UTAMP to the case of MMV with the model
\begin{eqnarray}
\bY=\sum_{k=1}^K b_k \bA_k \bC +\bW \label{eq:MultiY}
\end{eqnarray}
where $\bY$ is an observation matrix with size $M \times L$, $\bW$ denotes  a white Gaussian noise matrix with mean 0 and precision $\beta$, matrices $\{\bA_k\}$ are known, and $\bC$ with size $N \times L$ and $\bb=\left[b_1,...,b_K\right]^T$ are to be estimated.

Similar to the case of SMV, (\ref{eq:MultiY}) can be reformulated as
\begin{eqnarray}
\bY=\bA\bX+\bW \label{eq:MultiY_Linear}
\end{eqnarray}
where $\bA=[\bA_1, ...,\bA_K]$, and $\bX=[\bx_1,...,\bx_L]$ with
\begin{eqnarray}
\bx_l= \bb\otimes \bc_l. \label{eq:x_l}
\end{eqnarray}
With the SVD $\bA= \bU\bLambda\bV$ and unitary transformation, we have the following model
\begin{eqnarray}
\bR=\bPhi\bX+\overline{\bW} \label{eq:MultiY_Linear2}
\end{eqnarray}
where $\bR=\bU^{H}\bY$, $\bPhi=\bLambda\bV=\bU^H\bA$ and $\overline{\bW}=\bU^{H}\bW$. Define $\bz_l=\bm{\Phi}\bx_l$ and $\bZ=[\bz_1, ..., \bz_L]$, then we can factorize the joint distribution of the variables in \eqref{eq:MultiY_Linear2} as
\begin{eqnarray}
&&\!\!\!\!\!p(\bX,\bC,\bb,\bZ,\beta |\bR)\nonumber\\
&&\!\!\!\!\! \propto p(\bC)p(\bb)p(\beta)\prod\nolimits_{l}p(\br_l|\bz_l,\beta)p(\bz_l|\bx_l)p(\bx_l|\bb,\bc_l) \nonumber \\
&&\rv{\!\!\!\!\!\triangleq f_{\bC}(\bC)f_{\bb}(\bb) f_{\beta}(\beta)\prod\nolimits_{l}  f_{\br_l}(\bz_l,\beta)f_{\bz_l}(\bz_l,\bx_l){f_{\bx_l}(\bx_l,\bb,\bc_l)}}. \nonumber \\
\label{eq:FactorMulti}
\end{eqnarray}	

\begin{figure}[!t]
	\centering
	\includegraphics[width=0.5\textwidth]{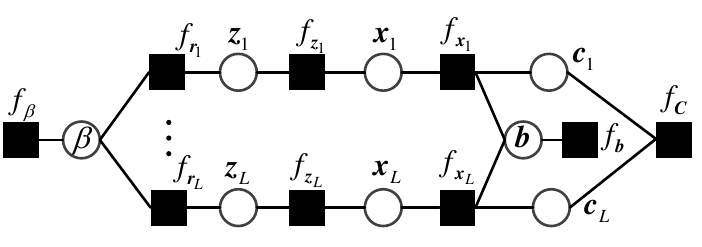}
	\centering
	\caption{\rv{Factor graph representation of \eqref{eq:FactorMulti}.}}
	\label{fig:ModelMMV}
\end{figure}

\begin{algorithm}
	\caption{Bi-UTAMP for MMV}
	\textbf{Unitary transform}: $\bR=\bU^H \bY =\bPhi \bX+\overline\bW$, where $\bA_{M \times NK}=\bU \bLambda \bV$, $\bPhi=\bU^H \bA=\bLambda \bV$, and $\bx_l = \bb\otimes \bc_l$ with $\bb=[b_1, ..., b_K]^T$ and $\bc_l=[c_{1,l}, ..., c_{N,l}]^T$. \\
    Let $\bPhi=\left[\bPhi_1, ..., \bPhi_K \right]$,  $\bphi_k=|\bPhi_k|^2 \textbf{1}_N$, and $\bx_l=\left[\bx_{1,l}^T, ..., \bx_{K,l}^T\right]^T,  k=1, ..., K, n=1, ..., N$ and $l=1, ..., L$. \\
    \textbf{Initialize}: $\hat b_{k}$, $\nu_{b_{k}}=1$, $\nu_{\bx_{k,l}}=1$, $\hat\bx_{k,l}=\textbf{0}$, $\mathbf{s}_l=\mathbf{ 0 }$, and $\hat \beta=1$. \\
	\textbf{Repeat}
	
	\begin{algorithmic}[1]
		\STATE $\forall l$: $\bnu_{\bp_l}=\sum_k\bphi_k\nu_{\bx_{k,l}}$
		\STATE $\forall l$: $\bp_l=\sum_k\bPhi_k\hat\bx_{k,l}-\bnu_{\bp_l}\cdot\bs_l$
		
		\STATE $\forall l$: $\bnu_{\bz_l}= \bnu_{\bp_l}./(\bm{1}+\hat\beta \bnu_{\bp_l})$
		\STATE $\forall l$: $\hat \bz_l=(\hat\beta\bnu_{\bp_l}\cdot \bm{r}_l+\bm{p}_l)./(\bm{1}+\hat\beta\bnu_{\bp_l})$
		
		\STATE $\hat\beta=ML/\sum\nolimits_l\big(\big\|\br_l-\hat \bz_l\big\|^2+ \textbf{1}^T \bnu_{\bz_l}\big)$
		\STATE $\forall l$: $\bnu_{\bs_l}=\bm{1}./\big(\bnu_{\bp_l}+\hat\beta^{-1}\textbf{1}_{M}\big)$
		\STATE $\forall l$: $\bs_l=\bnu_{\bs_l}\cdot\big(\br_l-\bp_l\big)$ \\
		
		\STATE $\forall l,k$: $\nu_{\bq_{k,l}}=1/\big<|\bPhi^H_k|^2\bnu_{\bs_l}\big>$
		\STATE $\forall l,k$: $\bq_{k,l}=\hat \bx_{k,l}+\nu_{\bq_{k,l}}\bPhi^H_k \bs_l$

		\STATE$\forall l,k:\vec \bc_{k,l}={\bq_{k,l}\hat b_{k}^*} /({|\hat b_{k}|^2+\nu_{b_{k}}})$ \\
		\STATE$\forall l,k:\vec\bnu_{\bc_{k,l}}=\textbf{1}_N{\nu_{\bq_{k,l}}}/ ({|\hat b_{k}|^2+\nu_{b_{k}}})$ \\
		\STATE$\forall l: \vec \bnu_{\bc_l} =\textbf{1}_N./ \sum_k (\textbf{1}_N./\vec \bnu _{\bc_{k,l}})$ \\
		\STATE$\forall l:\vec{\bc}_{l}=\vec\bnu_{\bc_{l}}\cdot\sum_k(\vec \bc_{k,l}./\vec\bnu_{\bc_{k,l}})$ \\
		\STATE$\forall n,l:\hat c_{n,l}=\mathbb{E}[c_{n,l}|\{\vec\bnu_{\bc_l},\vec \bc_l\}, f_{\bC}]$ \\
		\STATE$\forall n,l:\nu_{c_{n,l}}=\mathbb{V}\text{ar}[c_{n,l}|\{\vec\bnu_{\bc_l},\vec \bc_l\}, f_{\bC}]$ \\
	    \STATE$\forall l:\bnu_{\bc_l}=<[\nu_{c_{1,l}},...,\nu_{c_{N,l}}]>\textbf{1}_N, \hat \bc_l=[\hat c_{1,l}, ...,\hat c_{N,l} ]^T.$
	
		\STATE $\forall l,k: \vec\bnu_{\bb_{k,l}}={\nu_{\bq_{k,l}}} \textbf{1}_N./ ({\left|\hat \bc_{l}\right|^2+\bnu_{\bc_{l}}})$ \\
		\STATE$\forall l, k: \vec \bb_{k,l}={\bq_{k,l}\cdot \hat \bc_{l}^*}. /({\left|\hat \bc_{l}\right|^2+\bnu_{c_{l}}})$ \\
		\STATE$\forall k: \vec\nu_{b_k}=1/\sum_l(\textbf{1}_N^T (\textbf{1}./\vec\bnu_{\bb_{k,l}}))$
		\STATE$\forall k: \vec b_k=\vec\nu_{b_k} \sum_l(\textbf{1}_N^T (\hat \bb_{k,l}./\vec\bnu_{\bb_{k,l}}))$
        \STATE$\forall k:\hat b_k=\mathbb{E}[b_k|\{\vec\nu_{b_k},\vec b_k\}, f_{\bb}] $ \\
        \STATE$\forall k:\nu_{b_k}=\mathbb{V}\text{ar}[b_k|\{\vec\nu_{b_k},\vec b_k\}, f_{\bb}]$ \\

		\STATE$\forall l, k: \cev\bnu_{\bb_{k,l}}=\big(\nu_{b_k} \vec\bnu_{\bb_{k,l}}\big)./\big( \vec\bnu_{\bb_{k,l}}-\nu_{b_k} \textbf{1}_N\big) $\\
		\STATE$ \forall l, k:\cev\bb_{k,l}\!=\!\big(\hat b_k \vec\bnu_{ \bb_{k,l}}-\nu_{b_k} \vec\bb_k\ \big)./\big( \vec\bnu_{ \bb_{k,l}}-\nu_{b_k} \textbf{1}_N\big)$\\
		
		\STATE $\forall l, k$: $\cev\bnu_{\bc_{k,l}}=\left(\textbf{1}_N./\bnu_{\bc_l}-\textbf{1}_N./\vec\bnu_{\bc_{k,l}}\right)^{.-1} $
		\STATE $\forall l,k$: $\cev\bc_{k,l}=\cev\bnu_{\bc_{k,l}}\cdot\left(\hat\bc_l./\bnu_{\bc_l}-\vec\bc_{k,l}./\vec\bnu_{\bc_{k,l}}\right)$
		
		\STATE$\forall l, k$: $\cev \bx_{k,l}=\cev\bb_{k,l}\cdot\cev\bc_{k,l}$
		\STATE $\forall k,l$: $\cev\bnu_{ \bx_{k,l}}=|\cev\bb_{k,l}|^2\cdot\cev\bnu_{\bc_{k,l}}+\cev\bnu_{\bb_{k,l}}\cdot\left|\cev\bc_{k,l}\right|^2+
		\cev\bnu_{\bb_{k,l}}\cdot\cev\bnu_{\bc_{k,l}}$\\
		
		\STATE $\forall l, k$: $\bnu_{\bx_{k,l}}=\big(1/\nu_{\bq_{k,l}}\textbf{1}_N+\textbf{1}_N./\bnu_{\cev\bx_{k,l}}\big)^{.-1}$
		\STATE $\forall k,l$: $\hat \bx_{k,l}=\bnu_{\bx_{k,l}}\cdot\left(1/\nu_{\bq_{k,l}} \bq_{k,l}+\cev\bx_{k,l}./\cev\bnu_{\bx_{k,l}}\right)$
	    \STATE $\forall l, k: \nu_{\bx_{k,l}}= <\bnu_{\bx_{k,l}}> $
	\end{algorithmic}
	\textbf{Until terminated}
	\label{UTAMPMMV}
\end{algorithm}

The factor graph representation for the factorization in \eqref{eq:FactorMulti} is depicted in Fig. \ref{fig:ModelMMV}. The message updates related to $\bz_l$, $\bx_l$ and $\bc_l$ are the same as those in \textbf{Algorithm} \ref{BiUTAMPSMV}, and they can be computed in parallel. The major difference lies in the computations of $b(\bb)$ and $b(\beta)$, where the messages from $f_{\bx_l}$ and $f_{\br_l}, \forall l$, should be considered, i.e.,
\begin{eqnarray}
b(\bb)\propto \prod\nolimits_{l}m_{f_{\bx_l} \to \bb}(\bb)m_{f_{\bb} \to \bb}(\bb)\\
b(\beta)\propto \prod\nolimits_{l}m_{f_{\br_l} \to \beta}(\beta)m_{f_{\beta} \to \lambda}(\beta).
\end{eqnarray}
Similar to the SMV case, the message passing algorithm can be derived, which are summarized as \textbf{Algorithm} \ref{UTAMPMMV} (Bi-UTAMP for MMV).

\subsection{Discussions and Complexity Analysis}

We have the following remarks and discussions about Bi-UTAMP:

\begin{enumerate}
\item [1:] 	In some problems, $b_1$ is known, e.g., $b_1=1$.  In this case, we can set $\hat b_1=1$ and $\nu_{b_1}=0$ in Bi-UTAMP, which are indicated in \textbf{Algorithm} \ref{BiUTAMPSMV}.
\item [2:]  It is not hard to show that, when $\bb=b_1=1$, Bi-UTAMP is reduced to UTAMP (\textbf{Algorithm} \ref{UTAMP}) exactly.
\item [3:]  \rv{It is interesting that} the robustness of Bi-UTAMP can be enhanced by simply damping $\bs$, i.e.,  Line 7 of the SMV Bi-UTAMP is changed as
\begin{eqnarray}
\bs=(1-\alpha) \bs+ \alpha\bnu_{\bs}\cdot(\br-\bp)
\end{eqnarray}
with $\alpha\in(0,1]$, where $\alpha$ is the damping factor and $\alpha=1$ leads to the case without damping. Accordingly, Line 7 of the MMV Bi-UTAMP is changed as  $\bs_l=(1-\alpha) \bs_l+ \alpha\bnu_{\bs_l}\cdot(\br-\bp)$.
\item [4:] The iterative process can be terminated based on some criterion, e.g., the normalized difference between the estimates of $\bb$ of two consecutive iterations is smaller than a threshold, i.e., $\|\hat \bb^t-\hat\bb^{t-1}\|^2 /\|\hat \bb(t)\|^2 < \epsilon$ where  $\hat \bb^t$ is the estimate of $\bb$ at the $t$th iteration and $\epsilon$ is a threshold.
\item [5:] As the bilinear problem has local minima, we can use the same strategy of restart as in \cite{Sarkar2019} to mitigate the issue of being stuck at local minima. For each restart, we initialize $\{\hat b_k\}$ with different values.
\item [6:] \rv{In Bi-UTAMP, we have Bayesian treatment to both $\bb$ and $\bc$ (or $\bC$ in MMV). In contrast, $\bb$ is treated as a unknown deterministic variable in BAd-VAMP, and only a point estimate is involved. As discussed in Section III.A, the Bayesian treatment to $\bb$ can make the algorithm more flexible.}
\item [7:] The computational complexity of Bi-UTAMP is analyzed in the following. Bi-UTAMP needs pre-processing, i.e., performing economic SVD for $\bA$ and unitary transformation, and the complexity is $\mathcal O(M^2NK)$. It is noted that the pre-processing can be carried out offline (although we do not assume this \rv{in counting the runtime of Bi-UTAMP} in the simulations in Section IV). It can be seen from the Bi-UTAMP algorithms that, there is no matrix inversion involved, \rv{and the most computational intensive parts only involve matrix-vector products}. So the complexity of Bi-UTAMP per iteration is  $\mathcal O(MNKL)$ (in the case of SMV, $L=1$), which linearly increases with $M$, $N$, $K$ and $L$. \rv{For comparison, BAd-VAMP involves one outer loop and two inner loops. The whole matrix  $\bC_l^t$ with size $N \times N$ in the second inner loop is required in multiple lines in the algorithm and $\bA(\theta_{\bA}^t)$ is updated in each inner iteration \cite{Sarkar2019}. The computation of the matrix $\bC_l^t$ leads to a complexity of $\mathcal O(LN^3+KMN)$ per inner iteration. Line 18 is also computational intensive, which requires a complexity of $\mathcal O(K^2N^2)$ per inner iteration. Also, Line 20 of BAd-VAMP requires a complexity of $\mathcal O(K^3)$ per inner iteration. It is difficult to have a very precise complexity comparison analytically as the algorithms require different numbers of iterations to converge. So, in Section IV, we compare the runtime of several state-of-the-art algorithms as in \cite{Sarkar2019}.} As demonstrated in Section IV, with much less runtime, Bi-UTAMP can outperform the state-of-the-art algorithms significantly.
\end{enumerate}

\subsection{\rv{SE-Based Performance Prediction}}

\rv{From the derivation of Bi-UTAMP, we can see that Bi-UTAMP integrates VMP, BP, EP and UTAMP. The incorporation of UTAMP enables the approximate inference method to deal with the most computational intensive part with low complexity and high robustness. The rigorous performance analysis is difficult, but we make an attempt to predict its performance based on UTAMP SE heuristically. We track the output variance of the UTAMP module in the dash-dotted box with respect to the input variance.
However, the variances are about $\bx$ instead of $\bb$ and $\bc$ (or $\bC$ in the MMV case). The method is the same as the SE for (UT)AMP, i.e., we model $\bq_k = \bx_k + \bw_k$ as the input to the "denoiser" (which corresponds to $f_{\bx}$, $f_{\bb}$ and $f_{\bc}$ in the factor graph and involves EP and BP), where $\bw_k$ denotes a Gaussian noise with mean zero and variance $\tau_k$. However, it is difficult to find an analytic form for the output variance of the denoiser, which is also happened to (UT)AMP due to the priors. This can be solved by simulating the denoiser using $\bq_k = \bx_k + \bw_k$ with different variances of $\bw_k$ as input, so that a "function" in terms of a table can be established. In our case, besides the variance of $\bx$, the MSE of $\bb$ and $\bc$ can also be obtained as "byproduct", which allows us to predict the MSE of $\bb$ and $\bc$, while the variance of $\bx$ is used to determine $\tau_k$ analytically. As shown in Section IV, the prediction is fairly good in some cases. But, in some cases, it is not accurate. More accurate and rigorous performance analysis is our future work.
}

\section{Numerical Examples}

In this section, we evaluate the performance of Bi-UTAMP and compare it with the state-of-the-art \rv{bilinear recovery algorithms including the conventional non-message passing based algorithm WSS-TLS in \cite{Zhu2011CSMU}, and message passing based algorithms BAd-VAMP in \cite{Sarkar2019} and PC-VAMP in \cite{ZHU}. It is noted that PC-VAMP does not provide an estimate for $\bb$. Performance is evaluated in terms of} normalized MSE and runtime. Relevant performance bounds are also included for reference.

\begin{figure}[!t]
	\centering
	\includegraphics[width=0.5\textwidth]{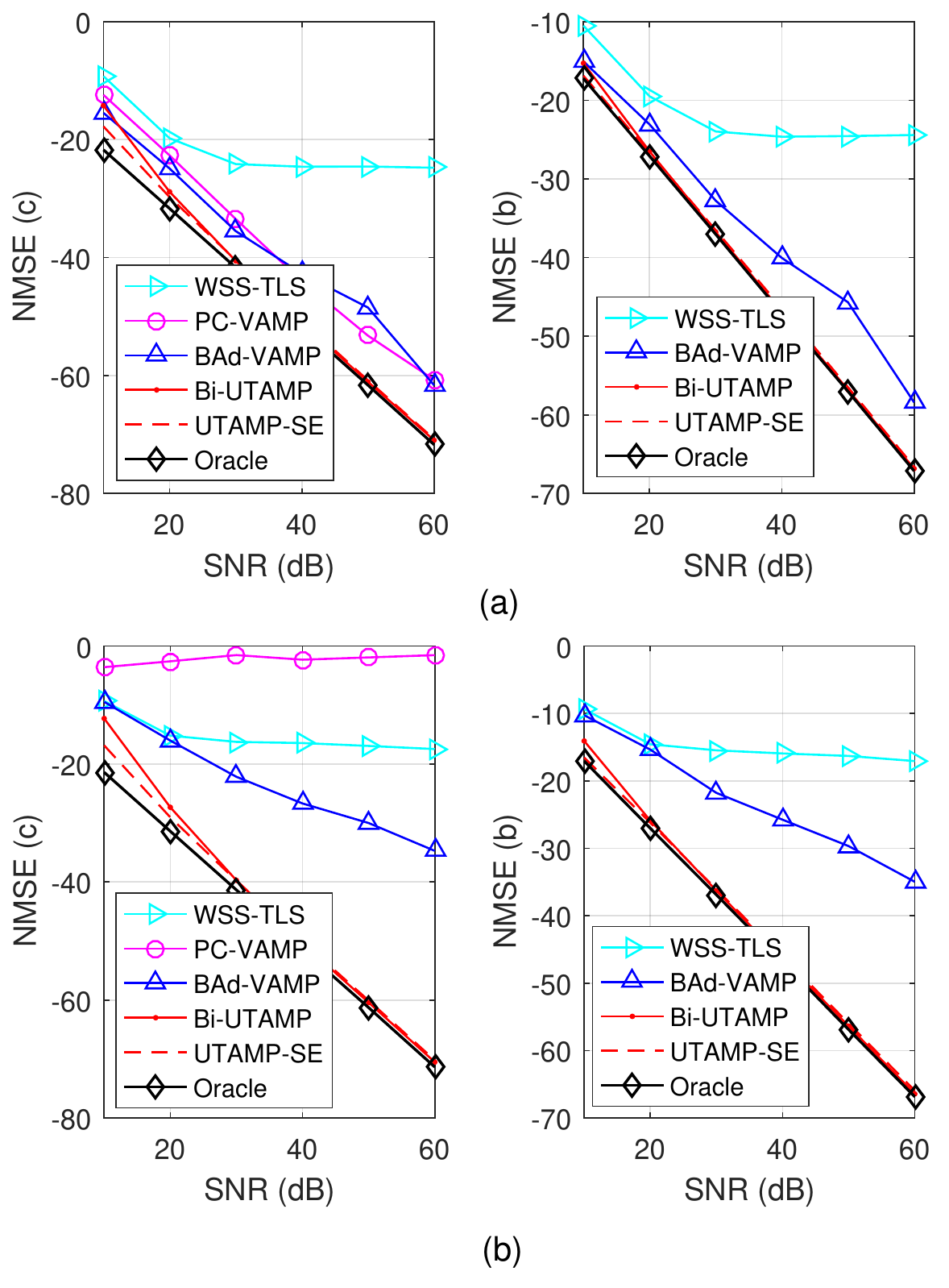}
	\centering
	\caption{\rv{Compressive sensing with correlated matrices: NMSE of $\bb$ and $\bc$ versus SNR with (a) $\rho=0.3$ and (b) $\rho=0.4$.}}
	\label{SMV_Corre_SNR}
\end{figure}

\subsection{SMV Case}

For the SMV case, we take compressive sensing with matrix uncertainty \cite{Zhu2011CSMU} as an example. We aim to recover a sparse signal vector $\bc$ from measurement $\by=\bA(\bb)\bc+\bw$, where the measurement matrix is modeled as $\bA(\bb)=\sum\nolimits_{k=1}^K b_k\bA_k$ with $b_1=1$, $\bA_k\in\mathbb{R}^{M\times N}$ are known, and the uncertainty parameter vector $\bb=[b_2, ..., b_K]^T$ is unknown. In addition the precision of the noise is unknown as well.

In the experiments, we set $K=11$, $N=256$, $M=150$ and the number of nonzero elements in $\bc$ is 10. The SNR is defined as $\text{SNR}\triangleq \mathbb{E}\left[||\bA(\bb)\bc||^2\right]/\mathbb{E}\left[||\bw||^2\right]$. The uncertainty parameters $\{b_2, ...b_k\}$ are drawn from $\N(0,1)$ independently, and the nonzero elements of sparse vector $\bc$ are drawn from $\N(0,1)$ independently as well, which are randomly located in $\bc$. The performance of the methods are evaluated using $\text{NMSE}(\bb)\triangleq ||\hat \bb-\bb||^2/||\bb||^2$ and $\text{NMSE}(\bc)\triangleq ||\hat \bc-\bc||^2/||\bc||^2$, where $\hat \bb$ and $\hat\bc$ are the estimates of $\bb$ and $\bc$, respectively. We also include the performance bounds for the estimation of $\bb$ and $\bc$, which are the performance of two oracle estimators: the MMSE estimator for $\bb$ with the assumption that $\bc$ is known, and the MMSE estimator for $\bc$ with the assumption that $\bb$ and the support of $\bc$ are known.

It is noted that, different from \cite{Sarkar2019}, we do not use median NMSEs, and \rv{to better evaluate the robustness of the algorithms}, the NMSEs are obtained by averaging the results from all trials. To demonstrate the robustness of Bi-UTAMP, we focus on tough measurement matrices, e.g., correlated matrices, non-zero mean matrices. and ill-conditioned matrices. In addition, Bi-UTAMP and BAd-VAMP use a same damping factor of 0.8 to enhance their robustness.

\begin{figure}[!t]
	\centering
	\includegraphics[width=0.5\textwidth]{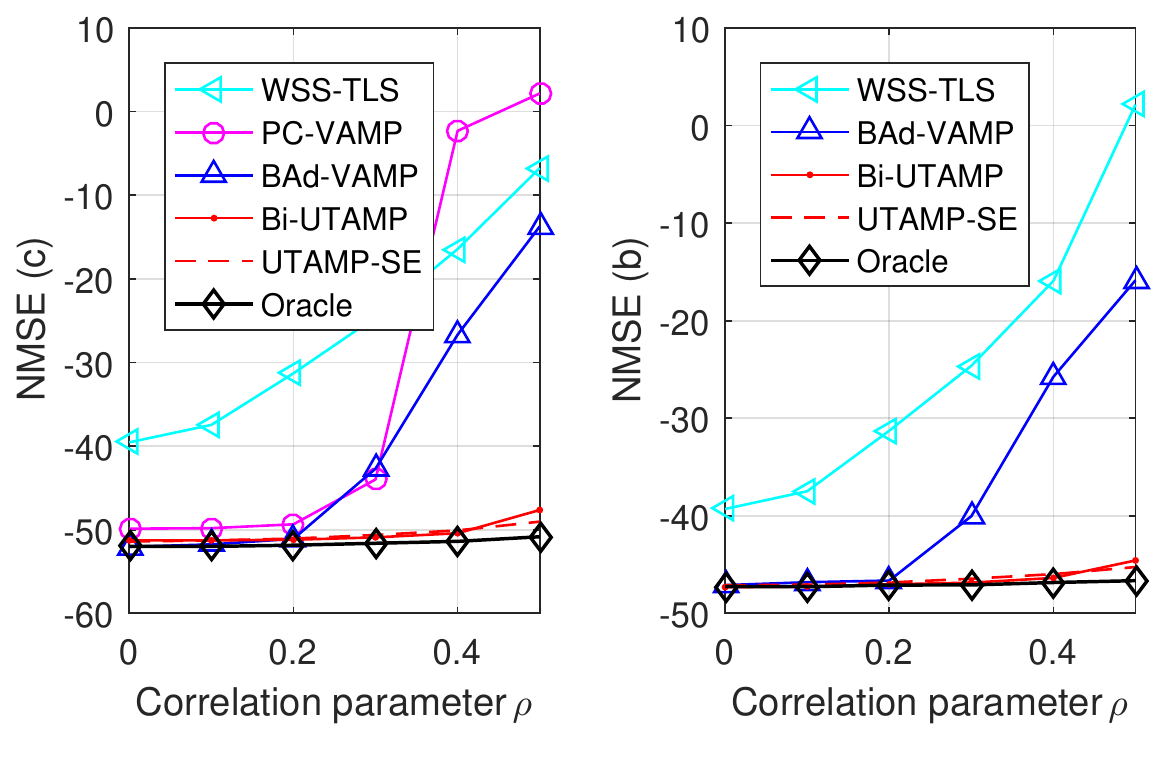}
	\centering
	\caption{\rv{Compressive sensing with correlated matrices: NMSE of $\bb$ and $\bc$ versus $\rho$ at SNR = 40dB.}}
	\label{fig:SMV_Corr}
\end{figure}

\subsubsection{Correlated Measurement Matrix}
All matrices $\{\bA_k\}$ are correlated, and $\bA_k$ is constructed using  $\bA_k=\bC_L\textbf{G}_k\bC_R$, where $\textbf{G}_k$ is an i.i.d. Gaussian matrix, and $\bC_L$ is an $M\times M$ matrix with the $(m,n)$th element given by $\rho^{|m-n|}$ where $\rho\in[0,1]$. Matrix $\bC_R$ is generated in the same way but with a size of $N\times N$. The parameter $\rho$ controls the correlation of matrix $\bA_k$. Fig. \ref{SMV_Corre_SNR} shows the NMSE performance of the algorithms versus SNR, where the correlation parameter $\rho=0.3$ in (a) and  $\rho=0.4$ in (b). It can be seen that when $\rho=0.3$, \rv{all the message passing based algorithms PC-VAMP, BAd-VAMP and Bi-UTAMP perform well and they are significantly better than the non-message passing based method WSS-TLS. We can also see that Bi-UTAMP delivers a performance which is considerably better than that of PC-VAMP and BAd-VAMP. With $\rho=0.4$, Bi-UTAMP still works very well, and it significantly outperforms BAd-VAMP, PC-VAMP and WSS-TLS. It is noted that as PC-VAMP does not estimate $\bb$, so its performance in the right column is absent.} We further evaluate the performance of \rv{all algorithms for matrices with different level of correlations by varying the parameter $\rho$} at SNR = 40dB and the results are shown in Fig. \ref{fig:SMV_Corr}, where we can see \rv{significant performance gaps between all the other algorithms and Bi-UTAMP when $\rho$ is relatively large. The results in Figs. \ref{SMV_Corre_SNR} and \ref{fig:SMV_Corr} demonstrate that Bi-UTAMP is  more robust than all the other algorithms with correlated measurement matrices. In Figs. \ref{SMV_Corre_SNR} and \ref{fig:SMV_Corr}, we also show the predicted performance based on SE for Bi-UTAMP, where we can see the the predicted performance matches the simulated performance fairly well when the matrix correlation is relatively small.}

\begin{figure}[!t]
	\centering
	\includegraphics[width=0.5\textwidth]{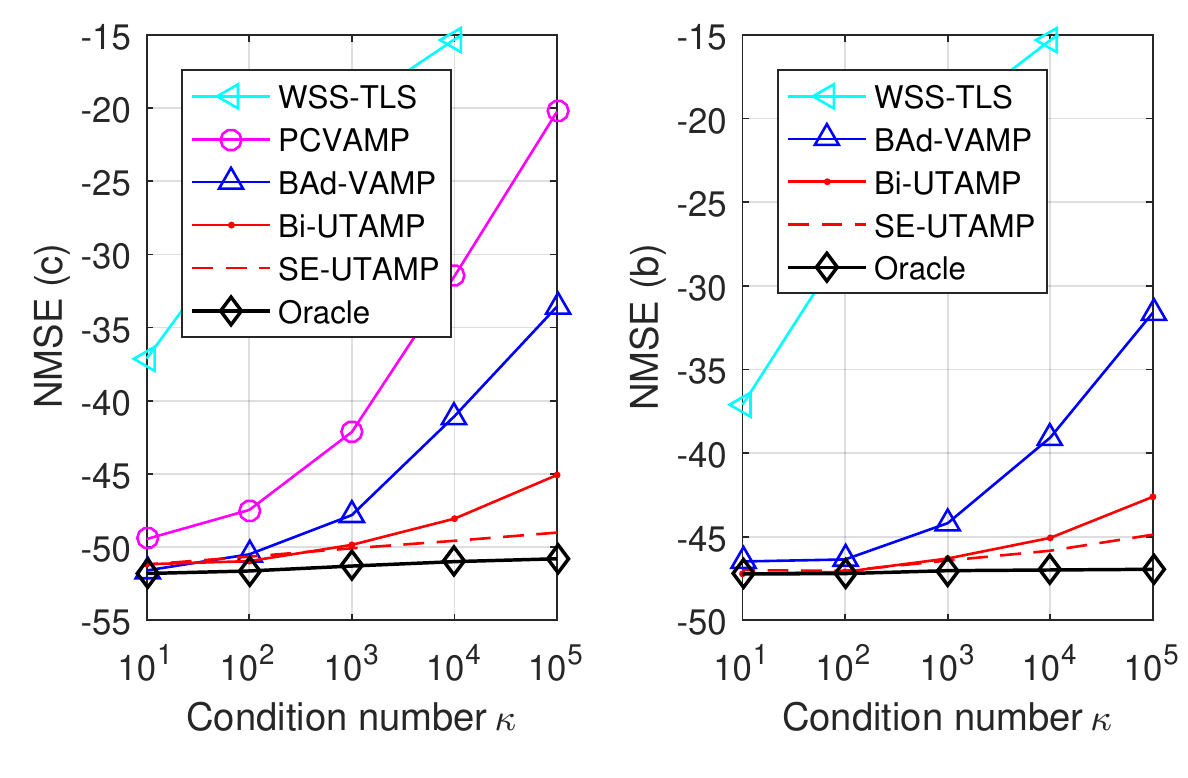}
	\centering
	\caption{\rv{Compressive sensing with ill-conditioned matrices: NMSE of $\bb$ and $\bc$ versus $\kappa$ with SNR = 40dB.}}
	\label{fig:SMV_Cond}
\end{figure}

\begin{figure}[!t]
	\centering
	\includegraphics[width=0.5\textwidth]{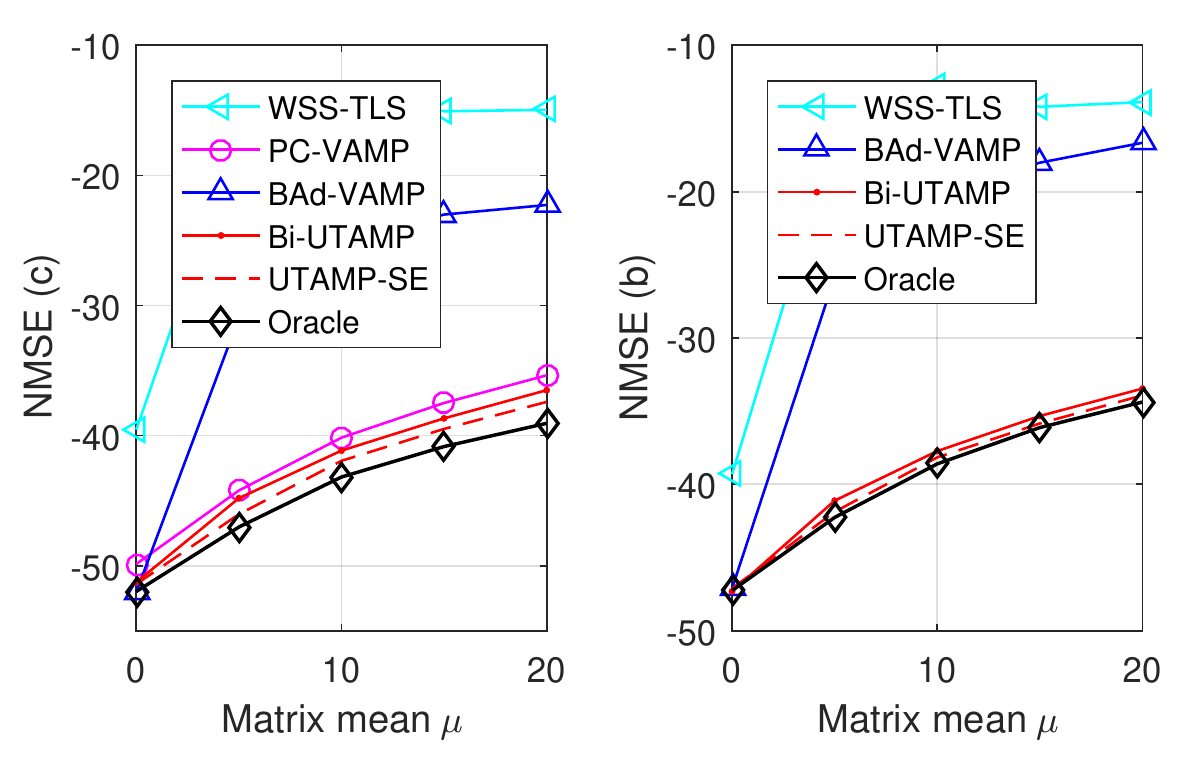}
	\centering
	\caption{\rv{Compressive sensing with non-zero mean matrix: NMSE of $\bb$ and $\bc$ versus $\mu$ with SNR = 40dB.}}
	\label{fig:SMV_Mu}
\end{figure}

\subsubsection{Ill-Conditioned Measurement Matrix}
Each matrix $\bA_k $ is constructed based on the SVD $\bA_k=\bU_k \bLambda_k \bV_k$ where $\bm{\Lambda}_k$ is a singular value matrix with $\Lambda_{i,i}/ \Lambda_{i+1,i+1} = \kappa^{1/(M-1)}$ (i.e., the condition number of the matrix is $\kappa$). The NMSE performance of the algorithms versus the condition number is shown in Fig. \ref{fig:SMV_Cond}, where the SNR = 40 dB. \rv{It can be seen that Bi-UTAMP can significantly outperform all the other algorithms when $\kappa$ is relatively large, and BAd-VAMP performs better than PC-VAMP and WSS-TLS. We also see that the predicated performance is no longer accurate when $\kappa$ is large.}

\subsubsection{Non-Zero Mean Measurement Matrix}
The elements of matrix $\bA_k$ are independently drawn from a non-zero mean Gaussian distribution $\mathcal{N}(\mu, v)$. The mean $\mu$ measures the derivation from the i. i. d. zero-mean Gaussian matrix. In the simulations,  for $\{\bA_k, k=2:K\}$, $v=1$, and for $\bA_1$, $v=20$. The NMSE performance of the algorithms versus $\mu$  is shown in Fig. \ref{fig:SMV_Mu}, where the SNR = 40 dB. It can be seen from this figure that Bi-UTAMP can achieve much better performance compared to \rv{WSS-TLS and BAd-VAMP especially when $\mu$ is relatively large. PC-VAMP delivers a competitive performance compared to Bi-UTAMP, while it does not provide an estimate for $\bb$ and is also slower than Bi-UTAMP as shown in Fig. \ref{fig:SMV_Time}.}

\subsubsection {Runtime Comparison}

\begin{figure}[!t]
	\centering
	\includegraphics[width=0.5\textwidth]{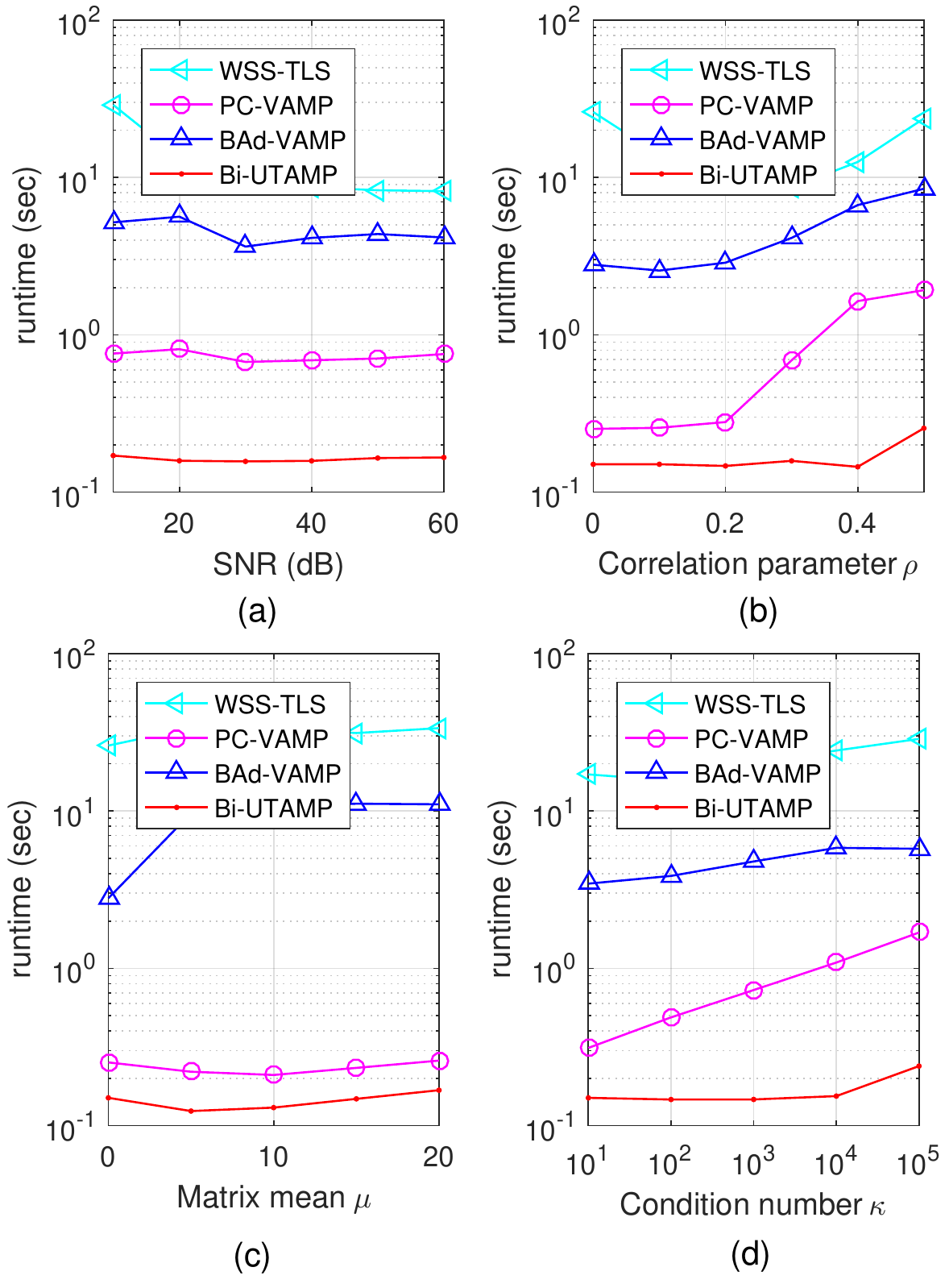}
	\centering
	\caption{\rv{Average runtime versus (a) SNR for correlated matrices with $\rho=0.3$, (b) $\rho$ for correlated matrices, (c) $\mu$ for non-zero mean matrices, (d) condition number $\kappa$ for ill-conditioned matrices. In (b), (c) and (d), SNR = 40 dB.}}
	\label{fig:SMV_Time}
\end{figure}

Fig. \ref{fig:SMV_Time} compares the average runtime of \rv{all algorithms.} In Fig. \ref{fig:SMV_Time} (a), correlated matrices are used with the correlation parameter $\rho=0.3 $. With SNR = 40 dB, the average runtime versus different $\rho$ for correlated matrices, different means for non-zero mean matrices and different condition numbers for ill-conditioned matrices is given in Fig. \ref{fig:SMV_Time} (b), (c) and (d), respectively. The results are obtained using MATLAB (R2016b) on a computer with a 6-core Intel i7 processor. Fig. \ref{fig:SMV_Time} shows that, Bi-UTAMP is much faster than BAd-VAMP \rv{and WSS-TLS, and it is also considerably faster than PC-VAMP.}

\begin{figure}[!t]
	\centering
	\includegraphics[width=0.5\textwidth]{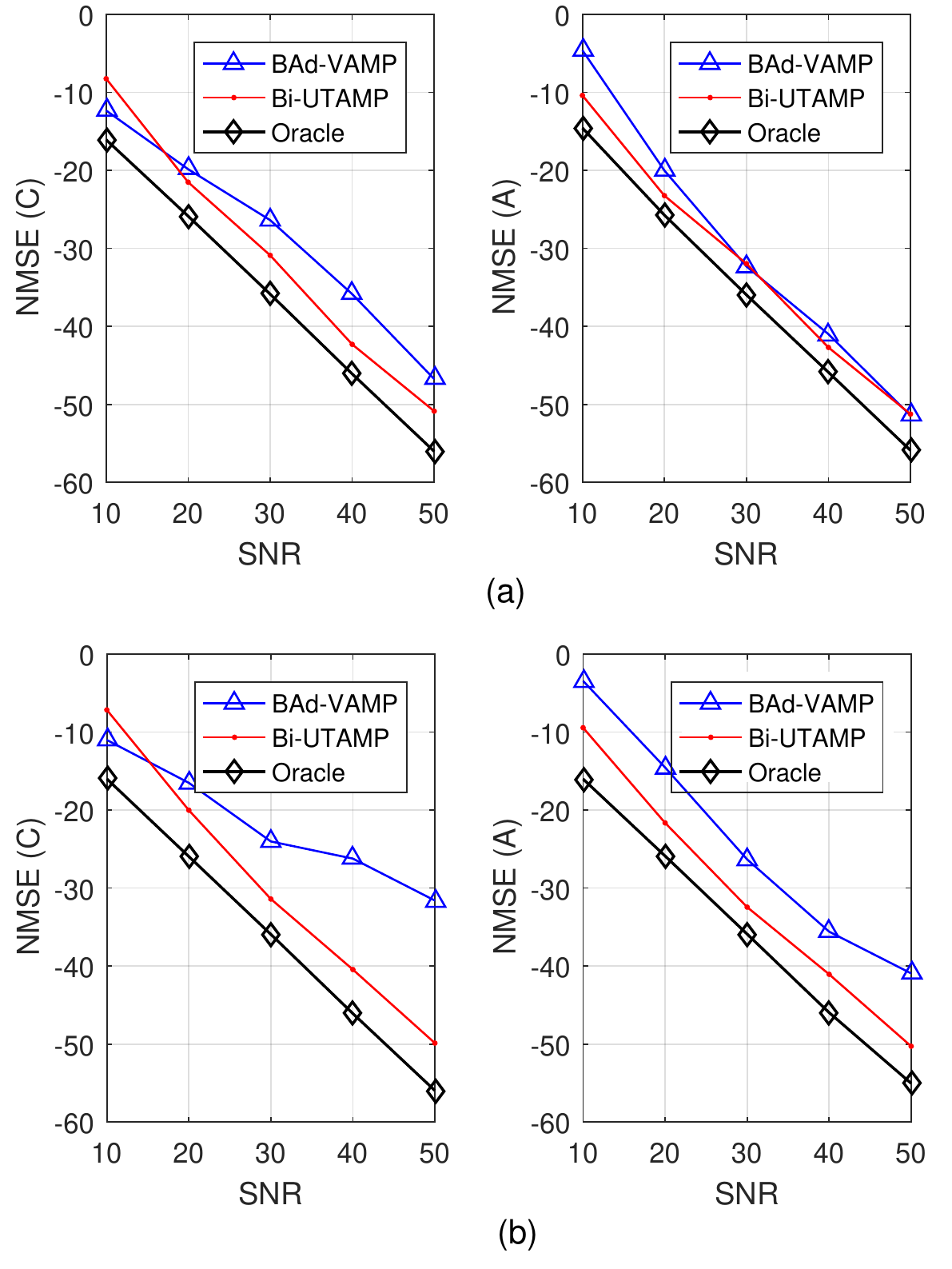}
	\centering
	\caption{Structured dictionary learning: NMSE($\bA$) and NMSE($\bC$) versus SNR with (a) $\rho$ = 0 and (b) $\rho$ = 0.1.}
	\label{fig:MMV_Snr}
\end{figure}


\subsection{MMV Case}

We take the structured dictionary learning (DL) \cite{Rubinstein2010DL} as an example to demonstrate the performance of Bi-UTAMP. The goal of structured DL is to find a structured dictionary matrix $\bA=\sum\nolimits_{k=1}^K b_k \bA_k\in\mathbb{R}^{M\times N}$ from the training samples $\bY\in\mathbb{R}^{M\times L}$ with model $\bY=\bA\bC+\bW$ for some sparse coefficient matrix $\bC\in\mathbb{R}^{N\times L}$. In the simulations, we assume square dictionary matrix $\bA$ with $M = N = 100$. The length of vector $\bb$ \rv{is large}, i.e., $K=100$, and the number of non-zero elements are set to be 20 in each column of $\bC$ \rv{(the columns are generated independently)} and $L=5$ for the training examples. Since the dictionary matrix $\bA$ has a structure, it can be learned with a small number of training samples. Bi-UTAMP is run for maximum $100$ iterations and $10$ restarts. In addition, to enhance the robustness, we use a damping factor $0.55$ for both Bi-UTAMP and BAd-VAMP. In addition, Lines 19-22 in Bi-UTAMP are executed once every two iterations. The performance is evaluated with NMSE of the estimates of $\bA$ and $\bC$. As the pair $(\bA, \bC)$ has a scalar ambiguity, the NMSE is calculated in the same way as in \cite{Sarkar2019}, i.e.,
\begin{equation}
\text{NMSE}(\hat\bA)\triangleq  \text{min}_{d}\frac{||\bA-d\hat\bA||^2}{||\bA||^2}\label{eq:NMMSE}
\end{equation}
\begin{equation}
\text{NMSE}(\hat\bC)\triangleq \text{min}_{d}\frac{||\bC-d\hat\bC||^2}{||\bC||^2}.
\end{equation}
Different from \cite{Sarkar2019}, the NMSEs are obtained by averaging the results from all trials. To test the performance and robustness of the algorithms, correlated matrices $\{\bA_k\}$ generated in the same way as in the SMV case are used.

\begin{figure}[!t]
	\centering
	\includegraphics[width=0.5\textwidth]{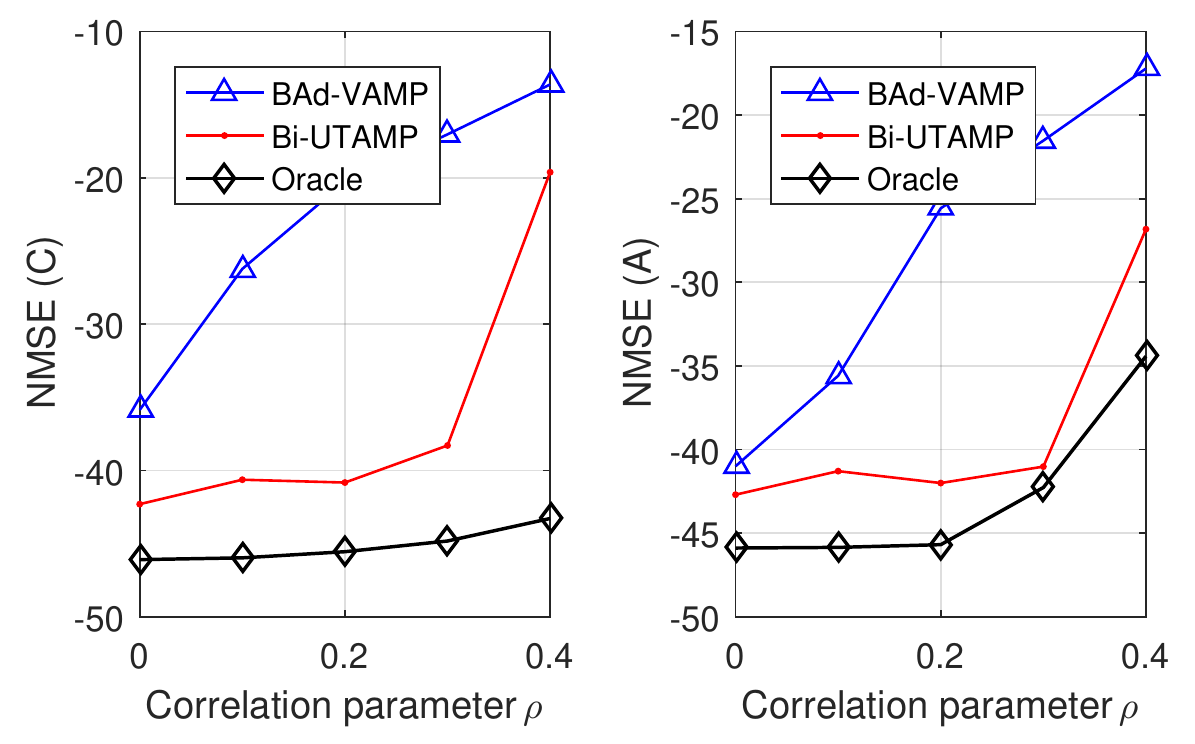}
	\centering
	\caption{Structured dictionary learning: NMSE(A) and NMSE(C) versus $\rho$ with SNR = 40dB.}
	\label{fig:MMV_Coef}
\end{figure}

\begin{figure}[!t]
	\centering
	\includegraphics[width=0.5\textwidth]{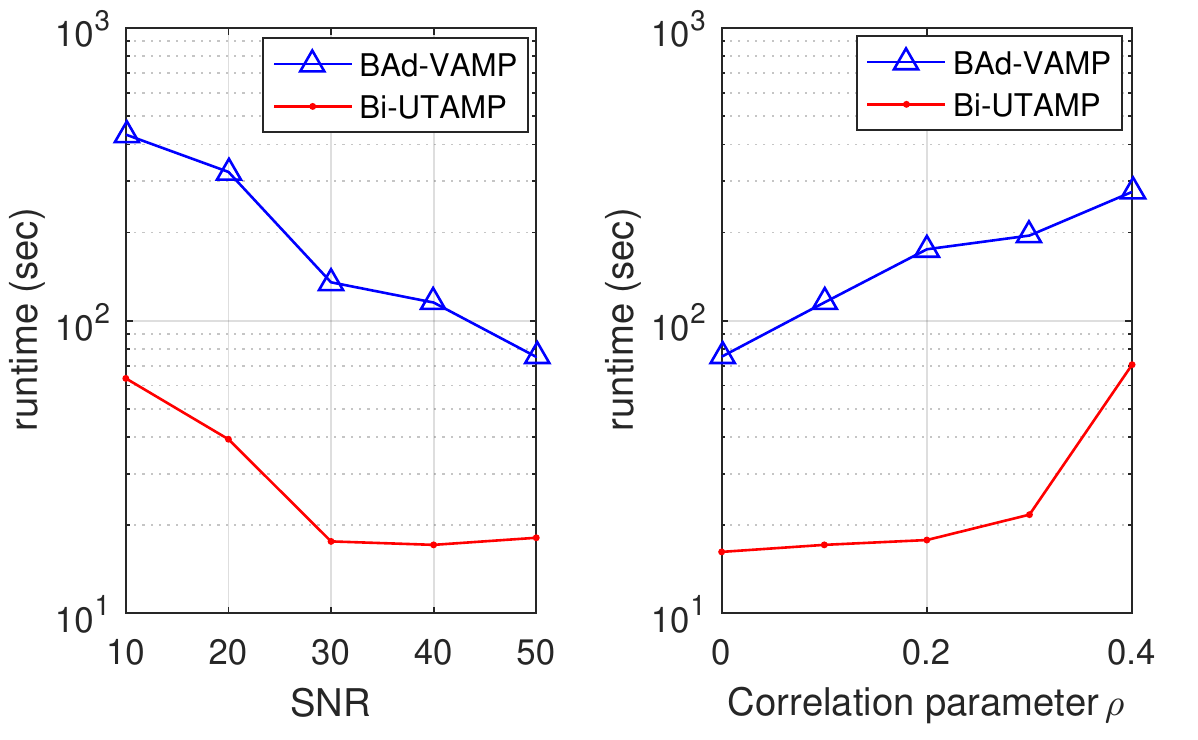}
	\centering
	\caption{{Structured dictionary learning: Average runtime versus SNR (left) and $\rho$ (right).}}
	\label{fig:MMV_TvsCoef}
\end{figure}

Figure \ref{fig:MMV_Snr} shows the NMSE performance $\text{NMSE}(\hat\bA)$ and $\text{NMSE}(\hat\bC)$ versus SNR with correlation parameter (a) $\rho$ = 0 and (b) $\rho$ = 0.1. It can be seen that when $\rho$ = 0, i.e., $\{\bA_k\}$ are i.i.d. Gaussian, BAd-VAMP and Bi-UTAMP have similar performance. When $\rho$ = 0.1, Bi-UTAMP  can outperform BAd-UTAMP considerably. Fig. \ref{fig:MMV_Coef} shows the NMSE versus $\rho$ at SNR = 40dB, where we can see that Bi-UTAMP can achieve significantly better performance than BAd-VAMP. From these results, we conclude that Bi-UTAMP is more robust. Figure \ref{fig:MMV_TvsCoef} shows the average runtime versus (a) SNR and (b) $\rho$. Again, the results show that Bi-UTAMP is much faster than BAd-VAMP.


\section{Conclusions}

In this paper, \rv{we have investigated approximate Bayesian inference for the problem of bilinear recovery. We have designed a new approximate inference algorithm Bi-UTAMP, where UTAMP is integrated with BP, EP and VMP} to achieve efficient recovery of the unknown variables. We have shown that Bi-UTAMP is much more robust and faster than the state-of-the-art algorithms, leading to significantly better performance. \rv{Future work includes a rigorous analysis of the performance of Bi-UTAMP and generalizing it to handle non-linear measurements, e.g., quantization.}

\section*{Acknowledgment}

The authors would like to thank Subrata Sarkar for sharing the Matlab code for BAd-VAMP and suggestions for the simulation of BAd-VAMP.

\bibliographystyle{IEEEtran}
\bibliography{IEEEabrv,bibliography}

\end{document}